\documentclass[11pt]{article}
\usepackage{amsmath,amsthm}
\usepackage{color,soul}
\usepackage{natbib}
\usepackage{hyperref}
\usepackage{booktabs}
\usepackage{amsfonts}
\usepackage{graphicx}
\usepackage[export]{adjustbox}
\graphicspath{ {figures/} }
\usepackage{cleveref}
\usepackage{pdflscape}
\usepackage{multirow}
\usepackage[small,bf]{caption}
\usepackage{subcaption}
\usepackage[stable]{footmisc}
\usepackage{bm}
\usepackage{booktabs}
\usepackage{lineno}

\usepackage{cancel}

\newtheorem{prop}{Proposition}[section]
\newenvironment{prop*}
  {\ex}
  {\endex}

\newtheorem{remark}{Remark}[section]
\newenvironment{remark*}
  {\ex}
  {\endex}

\newtheorem{definition}{Definition}[section]
\newenvironment{definition*}
  {\ex}
  {\endex}

\title{Measuring the frequency dynamics of financial connectedness and systemic risk\thanks{We thank the editor George Tauchen, an associate editor, two anonymous referees, and participants at various seminars and conferences for their comments, which have greatly improved the paper. Support from the Czech Science Foundation under the GA16-14179S project is gratefully acknowledged.}
\thanks{For estimating the frequency-dependent connectedness measures introduced by this paper, we provide the package \texttt{frequencyConnectedness} in \textsf{R} software. The package is available on CRAN or \url{https://github.com/tomaskrehlik/frequencyConnectedness}. Note that the term \textit{connectedness} is used widely in the literature studying how the variables in a system are connected \citep{diebold2011network}. In focusing on sources of systemic risk, we focus in particular on sources of system-wide connectedness.
}}

\author{%
Jozef {\sc Barun\'{i}k}$^{\rm a,b}$\thanks{Corresponding author, Tel. +420(776)259273, Email address: barunik@fsv.cuni.cz}, and
Tom\'{a}\v{s} {\sc K\v{r}ehl\'ik}$^{\rm a,b}$
\vspace{5mm} \\
 \small $^{\rm a}$ Institute of Economic Studies, Charles University, \vspace{-0.5mm}\\  \
 \small Opletalova 26, 110 00, Prague, Czech Republic \vspace{3mm} \\
 \small $^{\rm b}$ Department of Econometrics, IITA, The Czech Academy of Sciences, \vspace{-0.5mm}\\
 \small Pod Vodarenskou Vezi 4, 182 00, Prague, Czech Republic}

\begin{document}
\maketitle
\begin{abstract}
\noindent 
We propose a new framework for measuring connectedness among financial variables that arises due to heterogeneous frequency responses to shocks. To estimate connectedness in short-, medium-, and long-term financial cycles, we introduce a framework based on the spectral representation of variance decompositions. In an empirical application, we document the rich time-frequency dynamics of volatility connectedness in US financial institutions. Economically, periods in which connectedness is created at high frequencies are periods when stock markets seem to process information rapidly and calmly, and a shock to one asset in the system will have an impact mainly in the short term. When the connectedness is created at lower frequencies, it suggests that shocks are persistent and are being transmitted for longer periods. \\

 
\end{abstract}

\section{Introduction}
\label{sec:introduction}

The connectedness of financial markets is central to many areas of research, including risk management, portfolio allocation, and business cycle analysis. Being painfully aware of the unsuitability of standard correlation-based measures, academics have concentrated on developing more general frameworks. An abundant body of literature, however, still overlooks several fundamental properties of connectedness and hence possible sources of systemic risk. In our work, we argue that to understand the sources of connectedness in an economic system, it is crucial to understand the frequency dynamics of the connectedness, as shocks to economic activity impact variables at various frequencies with various strengths. To consider the long-, medium-, and short-term frequency responses to shocks, we propose a general framework that will allow us to measure the financial connectedness at a desired frequency band.

The main reason why we should believe that agents operate on different investment horizons represented by frequencies lies in the formation of their preferences. \cite{ortu2013long} disaggregate consumption growth into cyclical components classified by their level of persistence, and they develop an asset pricing model in which consumption responds to shocks due to heterogeneous preference choices. The authors hence extend the growing literature on consumption-based asset pricing models that price long-run risk in consumption growth \citep{bansal2004risks}. In an earlier contribution, \cite{cogley2001frequency} decomposes the approximation errors in stochastic discount factor models by frequency and applies the frequency decomposition to a number of consumption-based discount factor models. \cite{bandi2015business} further argue that consumption growth should be separated into a variety of cyclical components because investors may not focus on very high-frequency components of consumption representing short-term noise; instead, they may focus on lower-frequency components of consumption growth with heterogeneous periodicities. In a financial system, asset prices driven by consumption growth with different cyclical components will naturally generate shocks with heterogeneous frequency responses, and thus, various sources of connectedness will create short-term, medium-term, and long-term systemic risk. In turn, when studying connectedness, we should focus on linkages with various degrees of persistence underlying systemic risk.

The importance of the distinction between the short-term and the long-term parts of the system became evident even earlier with the dawn of co-integration \citep{engle1987co}. Subsequent literature builds a preliminary notion of disentangling short-term from long-term movements in connectedness \citep{gonzalo2001systematic,blanchard1989dynamic,quah1992relative}. Given the decomposition to the long-term common stochastic trend and deviations from the trend, one can move the projection in such a way that an error to one series will be a shock to the long-term trend, an error to another will be a shock to the deviation from the trend. A shock with a strong long-term effect will have high power at low frequencies, and in case it transmits to other variables, it points to long-term connectedness. For example, in the case of stock markets, long-term connectedness may be attributed to permanent changes in expectations about future dividends \citep{balke2002low}. To capture the frequency dynamics of connectedness, we propose a general framework for decomposing the connectedness to any frequency band of interest. Similar to \citet{dew2013asset}, who set asset pricing into the frequency domain, we view the frequency domain as a natural place for measuring the connectedness.

Focusing on shocks of an individual financial institution that have an impact on the wider system, our research contributes to the large body of literature measuring systemic risk both theoretically and empirically. As noted by \cite{benoit2016risks}, system-wide connectedness, or systemic risk, is often considered a ``hard-to-define-but-you-know-it-when-you-see-it'' concept. Generally, the literature views systemic risk as the risk that many market participants are simultaneously affected by severe losses, which then spread through the system.\footnote{For a comprehensive review of the systemic risk literature, see \cite{benoit2016risks}.} The financial crisis of 2007-2009 has reminded us that liquidity shocks, insolvency, and losses can quickly propagate and affect institutions, even in different markets. Consequently, there has been a growing demand for the design of financial regulations that are aimed to control individual institutions. From the perspective of market supervision, securing ourselves against this type of risk requires a sound quantification of systemic risk. Whereas measures specific to a particular risk channel are useful for calibrating regulatory tools, global measures aiming to quantify the contribution of financial institutions to total systemic risk are necessary to identify systematically important institutions. In cases where the contribution of institutions is persistent--instead of affecting the system solely in the short term--the systemically important financial institutions (SIFIs) may then be subject to higher capital requirements or a systemic risk tax. Because systemic risk threatens the stability of the entire financial sector, knowing the frequency-specific source of the instability is key for policymakers who are looking for tools to monitor the accumulation of risk.

Individuals who are interested in the frequency sources of connectedness in variables may consider using different forecast horizons of variance decomposition.\footnote{As noted by \cite{diebold2009,diebold2012better}, and later \cite{diebold2011network}, variance decompositions from approximating models are a convenient framework for empirical measurements of connectedness. More precisely, \cite{diebold2009} define the measures based on assessing shares of forecast error variation in one variable due to a shock arising in another variable in the system.} Staying in the time domain, heterogeneous frequency responses to shocks are simply aggregated through frequencies. To see this, let us consider two examples of a system of a bivariate autoregressive (AR) process with opposite signs of coefficients. The positive coefficients in the first example will create large connectedness driven by low frequencies of the cross-spectral density. With the increasing forecast horizon of variance decompositions, one will measure higher connectedness in the process. In the second example, the negative coefficients of the same magnitude will create equal connectedness as in the first case at all forecasting horizons, although connections come solely from the high frequencies due to the anti-persistent nature of the process. Hence, simply assessing connectedness at different horizons to capture the heterogeneous frequency responses due to differing expectations of investors is not sufficient.

Instead of assessing the overall error variation in a variable \emph{a} due to shock arising in a variable \emph{b}, we are interested in assessing shares of forecast error variation in a variable \emph{a} due to shock to a variable \emph{b} at a specific frequency band. This is a natural step to take, as it will show the long-term, medium-term, and short-term impacts of a shock, which can conveniently be summed to a total aggregate effect, if needed. For the purpose of frequency-dependent measurement, we define the spectral representation of generalized forecast error variance decomposition. To achieve this, we work with the Fourier transforms of the impulse response functions, i.e., frequency responses. In the frequency domain, we are simply interested in the portion of forecast error variance at a given frequency band that is attributed to shocks in another variable. Our work is inspired by the previous research of \cite{geweke1982measurement,geweke1984measures,geweke1986superneutrality} and \cite{stiassny1996spectral}, who use related measures in more restrictive environments.

In addition to introducing frequency dynamics into the measurement of connectedness, we study how cross-sectional correlations impact the connectedness. A higher contemporaneous correlation does not necessarily indicate connectedness in the sense that the literature tries to measure it. A good example is the recent crisis of 2007-2008, when stock markets recorded strong cross-sectional correlations that biased the contagion effects estimated by many researchers \citep{forbes2002no,bekaert2005market}.

The paper starts with a theoretical discussion. This is followed by a relevant application on financial data that show the usefulness of the framework and guide users in applying the introduced methods appropriately. Concretely, we study an important problem of connectedness in financial systems with a special focus on the frequency-specific measurement of systemic risk. We use the spectral representations of variance decompositions locally to recover the time-frequency dynamics of the connectedness of the main US financial institutions, and we document rich dynamics in the frequency responses of shocks in volatilities. We find that the dynamics of connectedness are not exclusively driven by one band of frequencies; different frequency bands play varying roles at different times. The dynamic corresponds intuitively to the events that occurred in the global financial markets.

\section{Measuring connectedness in frequency domain} 
\label{sec:methodology_sim_stud}

System connectedness can be characterized through variance decompositions from a vector auto-regression approximating model \citep{diebold2012better,diebold2009}. Variance decompositions provide useful information about how much of the future uncertainty of variable $i$ is due to shocks in variable $j$. One can measure how the system is interconnected using aggregation of the information in variance decompositions for many variables. \cite{diebold2011network} further argue that variance decompositions are closely linked to modern network theory as well as recently proposed measures of various types of systemic risk, such as expected shortfall \citep{acharya2017measuring} and CoVaR \citep{adrian2016covar}.

A natural way to describe the frequency dynamics (the long-term, medium-term, or short-term) of the connectedness is to consider the spectral representation of variance decompositions based on frequency responses to shocks. \citet{stiassny1996spectral} introduced a first notion of spectral representation for variance decompositions, albeit in a restrictive setting. In our work, we define the general spectral representation of variance decompositions, and we show how we can use it to define the frequency-dependent connectedness measures.

The spectral representations of variance decompositions can also be viewed as a more general way of measuring causality in the frequency domain. \cite{geweke1982measurement} proposes a frequency domain decomposition of the usual likelihood ratio test statistic for Granger causality, and \citet{dufour1998short,breitung2006testing,Yamada2014Some-Theoretica} provide a formal framework for testing causality on various frequencies. \cite{geweke1984measures,granger1969investigating} develop a multivariate extensions; however, all the analysis is done using partial cross-spectra and is therefore silent on the indirect causality chains. Hence, we are also motivated by this part of the econometrics literature to propose a more general framework.

Before defining the connectedness measures in the frequency domain, we briefly discuss the method of measuring connectedness introduced by \cite{diebold2012better} using generalized forecast error variance decompositions (GFEVD), as we build on these ideas in the frequency domain later in the text.

\subsection{Measuring connectedness with variance decompositions} 
\label{ssub:connectedness}

The connectedness measures are built from the variance decomposition matrix of a vector autoregressive (VAR) approximating model. In particular, consider a covariance stationary $N$-variate process $\mathbf{x}_t=(x_{1,t},\ldots,x_{N,t})'$ at $t=1,\ldots,T$ described by the VAR model of order $p$ as 
$$\mathbf{x}_t = \boldsymbol\Phi_1 \mathbf{x}_{t-1} + \boldsymbol\Phi_2 \mathbf{x}_{t-2} + \ldots + \boldsymbol \Phi_p \mathbf{x}_{t-p} + \boldsymbol \epsilon_t,$$ 
with $ \boldsymbol \Phi_1,\ldots, \boldsymbol \Phi_p$ coefficient matrices, and $\boldsymbol \epsilon_t$ being white noise with (possibly non-diagonal) covariance matrix $\boldsymbol \Sigma$. In this model, each variable is regressed on its own $p$ lags as well as the $p$ lags of each of the other variables in the system; hence, matrices of the coefficients contain complete information about the connections between variables.  It is useful to work with $(N \times N)$ matrix lag-polynomial $\boldsymbol \Phi(L)=[ \boldsymbol I_N - \boldsymbol \Phi_1 L - \ldots - \boldsymbol \Phi_p L^p]$ with $\boldsymbol I_N$ identity matrix, as the model can be written concisely as $\boldsymbol \Phi(L) \mathbf{x}_t = \boldsymbol \epsilon_t$. Assuming that the roots of $| \boldsymbol \Phi(z) |$ lie outside the unit circle, the VAR process has the following vector moving average (i.e., MA($\infty$)) representation $$\mathbf{x}_t = \boldsymbol \Psi(L) \boldsymbol \epsilon_t,$$ where $\boldsymbol \Psi(L)$ matrix of infinite lag polynomials can be calculated recursively from $\boldsymbol \Phi(L) = [\boldsymbol \Psi(L)]^{-1}$ and is key to understanding dynamics. Since $\boldsymbol \Psi(L)$ contains an infinite number of lags, it needs to be approximated with the moving average coefficients $\boldsymbol \Psi_h$ calculated at $h=1,\ldots,H$ horizons. The connectedness measures rely on variance decompositions, which are transformations of the $\boldsymbol \Psi_h$ and allow the measurement of the contribution of shocks to the system. 

Since a shock to a variable in the model does not necessarily appear alone, \emph{i.e.}, orthogonally to shocks to other variables, an identification scheme is crucial step in the calculation of variance decompositions. Standard approaches relying on Cholesky factorization depend on the ordering of the variables and complicate the measures. Generalized identification proposed by \cite{pesaran1998generalized} produces variance decompositions invariant to ordering.

Generalized variance decompositions can be written in the form\footnote{Note to notation: $(\boldsymbol A)_{j,k}$ denotes the $j$th row and $k$th column of matrix $\boldsymbol A$ denoted in bold. $(\boldsymbol A)_{j,\cdot}$ denotes the full $j$th row; this is similar for the columns. A $\sum A$, where $A$ is a matrix that denotes the sum of all elements of the matrix $A$.} (for a detailed derivation of the formula, see Appendix~\ref{app:genfevd})
\begin{equation}
	\left( \boldsymbol \theta_H \right)_{j,k} = \frac{\sigma_{kk}^{-1} \sum_{h = 0}^{H}\left((\boldsymbol \Psi_h \boldsymbol \Sigma)_{j,k}\right)^2}{\sum_{h=0}^{H} (\boldsymbol \Psi_h \boldsymbol \Sigma \boldsymbol \Psi_h')_{j,j}},
	\label{eq:connectedness}
\end{equation} where $\boldsymbol \Psi_h$ is a $(N \times N)$ matrix of moving average coefficients at lag $h$ defined above, and $\sigma_{kk} = \left( \boldsymbol \Sigma \right)_{k,k}$. The $\left( \boldsymbol \theta_H \right)_{j,k}$ denotes the contribution of the $k$th variable to the variance of forecast error of the element $j$, at horizon $h$. As the rows of the variance decomposition matrix $\boldsymbol \theta_H$ do not necessarily sum to one, each entry is normalized by the row sum as
 $$\left(\widetilde{\boldsymbol \theta}_{H}\right)_{j,k} = \left( \boldsymbol \theta_{H}\right)_{j,k}/\sum_{k=1}^N \left( \boldsymbol \theta_{H}\right)_{j,k}.$$
 Now the $\sum_{j=1}^N \left(\widetilde{\boldsymbol \theta}_{H}\right)_{j,k}=1$ and the sum of all elements in $\widetilde{\boldsymbol \theta}_{H}$ is equal to $N$, by construction. 

Note that $\left(\widetilde{\boldsymbol \theta}_{H}\right)_{j,k}$ provides a measure of pairwise connectedness from $j$ to $i$ at horizon $H$. This information can be aggregated to measure the total connectedness of the system. The connectedness measure is then defined as the share of variance in the forecasts contributed by errors other than own errors or as the ratio of the sum of the off-diagonal elements to the sum of the entire matrix \citep{diebold2012better}
\begin{equation}
  \mathcal{C}_{H} = 100 \cdot \frac{\sum_{j \neq k} \left(\widetilde{\boldsymbol \theta}_H\right)_{j,k} }{\sum \widetilde{\boldsymbol \theta}_H } = 100 \cdot \left(1 - \frac{\text{Tr} \left\{ \widetilde{\boldsymbol \theta}_H \right\} }{\sum \widetilde{\boldsymbol \theta}_H } \right),
  \label{eq:overallspillover}
\end{equation}
where $\text{Tr} \left\{ \cdot \right\}$ is the trace operator, and the denominator signifies the sum of all elements of the $\widetilde{\boldsymbol \theta}_H$ matrix. Hence, the connectedness is the relative contribution to the forecast variances from the other variables in the system.

\subsection{Spectral representation for variance decompositions and connectedness measures} 
\label{ssub:forecast_error_variance_decomposition_in_spectra}

A natural way to describe the frequency dynamics (the long-term, medium-term, or short-term) of the connectedness is to consider the spectral representation of variance decompositions based on frequency responses to shocks instead of impulse responses to shocks. As a building block of the presented theory, we consider a frequency response function, $\boldsymbol \Psi(e^{-i \omega})=\sum_h e^{-i\omega h} \boldsymbol \Psi_h$, which can be obtained as a Fourier transform of the coefficients $\boldsymbol \Psi_h$, with $i=\sqrt{-1}$. The spectral density of $\mathbf{x}_t$ at frequency $\omega$ can then be conveniently defined as a Fourier transform of MA($\infty$) filtered series as
$$
\boldsymbol S_\mathbf{x}(\omega)=\sum_{h=-\infty}^{\infty} E(\mathbf{x}_t\mathbf{x}'_{t-h})e^{-i\omega h}=\boldsymbol \Psi(e^{-i \omega}) \boldsymbol \Sigma \boldsymbol \Psi'(e^{+i\omega})
$$
The power spectrum $\boldsymbol S_\mathbf{x}(\omega)$ is a key quantity for understanding frequency dynamics, since it describes how the variance of the $\mathbf{x}_t$ is distributed over the frequency components $\omega$. Using the spectral representation for covariance, \emph{i.e.}, $E(\mathbf{x}_t\mathbf{x}'_{t-h})=\int_{-\pi}^{\pi} \boldsymbol S_\mathbf{x}(\omega) e^{i \omega h }d \omega$, the following definition naturally introduces the frequency domain counterparts of variance decomposition.

\begin{definition}
The generalized causation spectrum over frequencies $\omega\in(-\pi,\pi)$ is defined as
$$\left(\bm{\mathfrak{f}}(\omega) \right)_{j,k} \equiv \frac{\sigma_{kk}^{-1} \left| \left( \boldsymbol \Psi \left( e^{-i \omega}  \right) \boldsymbol \Sigma \right)_{j,k} \right|^2 }{\left(\boldsymbol \Psi(e^{-i \omega}) \boldsymbol \Sigma \boldsymbol \Psi'(e^{+i\omega})\right)_{j,j}},$$
where $\boldsymbol \Psi(e^{-i \omega})=\sum_h e^{-i\omega h} \boldsymbol \Psi_h$ is the Fourier transform of the impulse response $\boldsymbol \Psi_h$.
\end{definition}

It is important to note that $\left(\bm{\mathfrak{f}}(\omega) \right)_{j,k}$ represents the portion of the spectrum of the $j$th variable at a given frequency $\omega$ due to shocks in the $k$th variable. We can interpret the quantity as within-frequency causation, as the denominator holds the spectrum of the $j$th variable (on-diagonal element of cross-spectral density of $\mathbf{x}_t$) at a given frequency $\omega$. To obtain a natural decomposition of variance decompositions to frequencies, we can weight the $(\bm{\mathfrak{f}}(\omega))_{j,k}$ by the frequency share of the variance of the $j$th variable. The weighting function can be defined as
$$\Gamma_j(\omega)=\frac{\left(\boldsymbol \Psi(e^{-i \omega}) \boldsymbol \Sigma \boldsymbol \Psi'(e^{+i\omega})\right)_{j,j}}{\frac{1}{2 \pi} \int_{-\pi}^{\pi} \left(\boldsymbol \Psi(e^{-i \lambda}) \boldsymbol \Sigma \boldsymbol \Psi'(e^{+i\lambda})\right)_{j,j} d \lambda},$$
and it represents the power of the $j$th variable at a given frequency, which sums through frequencies to a constant value of $2\pi$. Note that while the Fourier transform of the impulse response is generally a complex valued quantity, the generalized causation spectrum is the squared modulus of the weighted complex numbers, thus producing a real quantity.

The following proposition establishes spectral representation of the variance decomposition, and it is central to the development of the connectedness measures in the frequency domain.

\begin{prop}
	\label{prop:1}
	Suppose $\mathbf{x_t}$ is wide-sense stationary with $\sigma_{kk}^{-1} \sum_{h = 0}^{\infty}\left|(\boldsymbol \Psi_h \boldsymbol \Sigma)_{j,k}\right| < +\infty, \forall j,k.$ Then, $$\left( \boldsymbol \theta_{\infty} \right)_{j,k} = \frac{1}{2 \pi} \int_{-\pi}^{\pi} \Gamma_j(\omega) \left(\bm{\mathfrak{f}} (\omega) \right)_{j,k} d \omega.$$
\end{prop}
\begin{proof}
	See Appendix \ref{app:proofs}.
\end{proof}

Using the result in Proposition \ref{prop:1}, $\left( \boldsymbol \theta_{H}\right)_{j,k}$ at $H\rightarrow\infty$ can be viewed as the weighted average of the generalized causation spectrum $\left(\bm{\mathfrak{f}}(\omega) \right)_{j,k}$, which gives us the strength of the relationship on given frequency weighted by the power of the series on that frequency. The integral over admissible frequencies perfectly reconstructs the theoretical value of the original $\left( \boldsymbol \theta_{\infty} \right)_{j,k}$. The proposition not only is an important theoretical result but also reminds us that when measuring connectedness with $\left( \boldsymbol \theta_{H}\right)_{j,k}$ at $H\rightarrow\infty$ in the time domain, we are looking at information aggregated through frequencies and ignoring heterogeneous frequency responses to shocks. It is also important to note that effects over the entire range of frequencies influence $\left( \boldsymbol \theta_{\infty} \right)_{j,k}$.

In economic applications, we are usually interested in assessing short-, medium-, or long-term connectedness rather than connectedness at a single given frequency. Hence, to better follow the economic intuition, it is more convenient to work with frequency bands that we define as the amount of forecast error variance created on a convex set of frequencies. The quantity is then given by integrating only over the desired frequencies $\omega \in (a,b)$.

Formally, let us have a frequency band $d=(a,b): a,b \in (- \pi, \pi), a < b.$ The generalized variance decompositions on frequency band $d$ are defined as
\begin{equation}
\left(\boldsymbol \theta_d \right)_{j,k} = \frac{1}{2 \pi} \int_d \Gamma_j(\omega) \left(\bm{\mathfrak{f}} (\omega) \right)_{j,k} d \omega.
\label{eq:thetaD}
\end{equation}

Because the introduced relationship is an identity and the integral is a linear operator, summing over disjoint intervals covering the entire range $(-\pi, \pi)$ will recover the original variance decomposition. The following remark formalizes this fact.

\begin{remark}
	\label{rem:recomposition}
	Denote by $d_s$ an interval on the real line from the set of intervals $D$ that form a partition of the interval $(-\pi, \pi)$, such that $\cap_{d_s \in D} d_s = \emptyset, $ and $\cup_{d_s \in D} d_s = (-\pi, \pi)$. Due to the linearity of integral and the construction of $d_s$, we have $$\left( \boldsymbol \theta_{\infty} \right)_{j,k} = \sum_{d_s \in D} \left( \boldsymbol \theta_{d_s} \right)_{j,k}.$$
\end{remark}

Using the spectral representation of generalized variance decomposition, it is straightforward to define connectedness measures on a given frequency band.

\begin{definition}
\label{def:freqmeasures}
	Let us define scaled generalized variance decomposition on the frequency band $d = (a,b): a,b \in (-\pi, \pi), a < b$ as $$\left( \widetilde{\boldsymbol \theta}_d \right)_{j,k} = \left( \boldsymbol \theta_d \right)_{j,k}/\sum_{k} \left( \boldsymbol \theta_{\infty} \right)_{j,k},$$ where $\boldsymbol \theta_d$ and $\boldsymbol \theta_\infty$ are defined as by Equation~(\ref{eq:thetaD}) and Proposition~\ref{prop:1}
	\begin{itemize}
    \item The \textbf{within connectedness} on the frequency band $d$ is then defined as $$\mathcal{C}^{\mathcal{W}}_d = 100 \cdot \left(1 - \frac{\text{Tr} \left\{  \widetilde{\boldsymbol \theta}_d \right\}}{\sum  \widetilde{\boldsymbol \theta}_d }\right).$$
		\item The \textbf{frequency connectedness} on the frequency band $d$ is then defined as
		$$\mathcal{C}^{\mathcal{F}}_d = 100 \cdot \left(\frac{\sum  \widetilde{\boldsymbol \theta}_d }{\sum  \widetilde{\boldsymbol \theta}_{\infty}} - \frac{\text{Tr} \left\{\widetilde{\boldsymbol \theta}_d \right\}}{\sum  \widetilde{\boldsymbol \theta}_{\infty} }\right) = \mathcal{C}^{\mathcal{W}}_d \cdot \frac{\sum  \widetilde{\boldsymbol \theta}_d }{\sum \widetilde{\boldsymbol \theta}_{\infty} },$$ 
	\end{itemize}
	where $\text{Tr} \left\{ \cdot \right\}$ is the trace operator, and the $\sum \widetilde{\boldsymbol \theta}_d$ signifies the sum of all elements of the $\widetilde{\boldsymbol \theta}_d$ matrix.
\end{definition}

The Definition \ref{def:freqmeasures} works with two notions: the \textit{frequency connectedness} and the \textit{within connectedness}. The \textit{within connectedness} gives us the connectedness effect that occurs within the frequency band and is weighted by the power of the series on the given frequency band exclusively. However, the \textit{frequency connectedness} decomposes the overall connectedness defined in Equation~(\ref{eq:overallspillover}) into distinct parts that, when summed, give the original connectedness measure $\mathcal{C}_{\infty}$. The following proposition formalizes the notion of reconstruction of the overall connectedness.

\begin{prop}[Reconstruction of frequency connectedness]
	\label{prop:2}
	Denote by $d_s$ an interval on the real line from the set of intervals $D$ that form a partition of the interval $(-\pi, \pi)$, such that $\cap_{d_s \in D} d_s = \emptyset, $ and $\cup_{d_s \in D} d_s = (-\pi, \pi)$. We then have that \begin{equation}
		\mathcal{C}_{\infty} = \sum_{d_s \in D} \mathcal{C}^{\mathcal{F}}_{d_s},
	\end{equation}
    where $\mathcal{C}_{\infty}$ is defined in Equation~(\ref{eq:overallspillover}) with $H \to \infty$.
\end{prop}
\begin{proof}
	See Appendix \ref{app:proofs}.
\end{proof}

To illustrate the difference between \emph{frequency} and \emph{within} connectedness, recall that the typical spectral shape of economic variables has the most power concentrated on low frequencies (long-term movements or trend). Hence, we could decompose the connectedness into two parts: one that covers long-term movements and another that covers short-term movements. Suppose that 90\% of the spectral density is concentrated in long-term movements, and hence, 10\% is in short-term movements. Now, suppose that the connectedness in short-term movements is high, say 80\%, and low on long-term movements, say 25\%. The 80\% and 25\% connectedness numbers represent the within connectedness. The total connectedness will be much closer to 25\% because the short-term connectedness of 80\% will be down-weighted by the very low amount (10\%) of spectral density on the short-term frequencies. Stated otherwise, although the short-term activities are very connected because of the small share of variance on the short-term frequencies, this connection becomes negligible in the aggregate system connectedness. This can be clearly observed in the simulations in the following section.

We conclude the theoretical section with the remark showing that the two concepts--the within and the frequency connectedness--coincide when the entire frequency band $d=(-\pi, \pi)$ is considered.

\begin{prop}
	\label{prop:3}
	Let us have $d=(-\pi, \pi)$. We then have
	\begin{equation}
		\mathcal{C}^{\mathcal{F}}_{d} = \mathcal{C}^{\mathcal{W}}_{d} = \mathcal{C}_{\infty}.
	\end{equation}
\end{prop}
\begin{proof}
	See Appendix \ref{app:proofs}.
\end{proof}

\subsection{Estimation of connectedness in the frequency domain}

The estimation of the previously defined connectedness measures relies heavily on the precise estimation of coefficients from the VAR approximating model. While standard VAR estimators work well in many cases, advanced techniques that include shrinkage or Bayesian approaches can help in situations of large dimensional data, deviations from distributional assumptions, etc.\footnote{Using precisely formulated time-varying parameter models could potentially allow the forecasting of connectedness measures.}

Variance decomposition of forecast errors is computed directly from moving average coefficients. Because the computation of these theoretical quantities is based on an infinite process, we make it feasible with a finite horizon $H$ approximation, as used in the definitions above, noting that the error from approximation disappears as $H$ grows (\cite{lutkepohl2007new}). The $\widehat{\boldsymbol \Psi}_h$ coefficients are then computed through standard recursive scheme $\widehat{\boldsymbol \Psi}_0 = \boldsymbol I, \widehat{\boldsymbol \Psi}_h = \sum_{j=1}^{\max \{h,p\}} \boldsymbol \Phi(j)\widehat{\boldsymbol \Psi}_{h-1},$ where $p$ is the order of VAR and $h \in \{1, \dots, H\}$. Here, we note that by studying the quantities in the frequency domain, $H$ serves only as an approximation factor, and it has no interpretation as in the time domain. In the applications, we advise setting the $H$ sufficiently high to obtain a better approximation, particularly when lower frequencies are of interest.

The spectral quantities are estimated using standard discrete Fourier transforms. The following definition accurately specifies the used estimates of the quantities.

\begin{definition}
\label{def:specest}
	The cross-spectral density on the interval $d = (a,b): a,b \in \left( -\pi, \pi \right), a < b$
	$$\int_{d} \boldsymbol \Psi(e^{-i \omega}) \boldsymbol \Sigma \boldsymbol \Psi'(e^{+i \omega}) d \omega$$
	 is estimated as
	$$\sum_{\omega} \widehat{\boldsymbol \Psi}(\omega) \widehat{\boldsymbol \Sigma} \widehat{\boldsymbol \Psi}'(\omega),$$
	for $\omega \in \left\{ \left\lfloor \frac{aH}{2 \pi} \right\rfloor, ..., \left\lfloor \frac{bH}{2 \pi} \right\rfloor \right\}$ where
	$$\widehat{\boldsymbol \Psi}(\omega) = \sum_{h=0}^{H-1} \widehat{\boldsymbol \Psi}_h e^{-2 i \pi \omega/H},$$ and $\widehat{\boldsymbol \Sigma} = \widehat{\boldsymbol \epsilon}'\widehat{\boldsymbol \epsilon}/(T-z)$, where $z$ is a correction for a loss of degrees of freedom, and it depends on the VAR specification.
\end{definition}

The decomposition of the impulse response function at the given frequency band is then estimated as $\widehat{\boldsymbol \Psi}(d) = \sum_{\omega} \widehat{\boldsymbol \Psi}(\omega)$. The definition \ref{def:specest} finally allows the estimation of the generalized variance decompositions at a desired frequency band as
$$\left( \widehat{\boldsymbol \theta}_d \right)_{j,k} = \sum_{\omega} \widehat{\Gamma}_j(\omega) \left( \widehat{\bm{\mathfrak{f}}} (\omega)\right)_{j,k},$$
where 
$$\left( \widehat{\bm{\mathfrak{f}}} (\omega)\right)_{j,k} \equiv \frac{\widehat{\sigma}_{kk}^{-1} \left( \left( \widehat{\boldsymbol \Psi}(\omega) \widehat{\boldsymbol \Sigma} \right)_{j,k} \right)^2}{\left( \widehat{\boldsymbol \Psi}(\omega) \widehat{\boldsymbol \Sigma} \widehat{\boldsymbol \Psi}'(\omega)\right)_{j,j}}$$
is estimated generalized causation spectrum, and 
$$\widehat{\Gamma}_j(\omega)=\frac{\left( \widehat{\boldsymbol \Psi}(\omega) \widehat{\boldsymbol \Sigma} \widehat{\boldsymbol \Psi}'(\omega)\right)_{j,j}}{\left( \Omega \right)_{j,j}},$$
is estimate of the weighting function, where $\Omega = \sum_{\omega} \widehat{\boldsymbol \Psi}(\omega) \widehat{\boldsymbol \Sigma} \widehat{\boldsymbol \Psi}'(\omega)$.

Then, the connectedness measures $\widehat{\mathcal{C}}^{\mathcal{W}}$ and $\widehat{\mathcal{C}}^{\mathcal{F}}$ at a given frequency band of interest can be readily derived by plugging the $\left( \widehat{\boldsymbol \theta}_d \right)_{j,k}$ estimate into the Definition~\ref{def:freqmeasures}.\footnote{The entire estimation is done using the package \texttt{frequencyConnectedness} in \textsf{R} software. The package is available on CRAN or on \url{https://github.com/tomaskrehlik/frequencyConnectedness}.}

\section{Simulation study} 
\label{sec:simulation_study}

To motivate the usefulness of the proposed measures, we study the processes that generate frequency-dependent connectedness by simulations. We look at connectedness that is induced through cross-sectional correlations or interactions between bivariate AR processes. The emergence of connectedness and its spectral footprints is then illustrated through a change in the coefficients in the bivariate VAR(1) case. Suppose the data have been generated from the following equations:
\begin{equation}
\label{eq:VARsimulation}
\begin{split}
	x_{1,t} = \beta_1 x_{1,t-1} + s x_{2,t-1} + \epsilon_{1,t}\\
	x_{2,t} = s x_{1,t-1} + \beta_2 x_{2,t-1} + \epsilon_{2,t},
\end{split}
\end{equation}
where $(\epsilon_{1,t}, \epsilon_{2,t}) \sim N(0, \boldsymbol \Sigma)$ with $\boldsymbol \Sigma = \begin{pmatrix}
	1 & \rho \\
	\rho & 1
\end{pmatrix}.$

By altering the true coefficients that generate the data, we study several cases with known values of theoretical connectedness estimates. We start with a symmetric process of $\beta = \beta_1 = \beta_2$, with three important cases generating distinctly connected variables $x_{1,t}$ and $x_{2,t}$. The first case is the $\beta = \beta_1 = \beta_2=0$, when we have two independent processes that have zero connectedness at all frequencies. Second, we study the connectedness of two symmetrically connected AR processes with the parameter $\beta = \beta_1 = \beta_2=0.9$ and $s=0.09$ or $\beta = \beta_1 = \beta_2=-0.9$ and $s=-0.09$ generating equal total connectedness with different sources from low and high frequencies of cross-spectral densities for positive and negative values of coefficients, respectively.

In addition to motivating the importance of the frequency dynamics of connectedness, we show the importance of cross-sectional correlations, which translates to all frequencies and may bias the connectedness measures. Hence, for all cases, we consider two extremes of cross-sectional dependence: no correlation $\rho=0$ and a correlation of $\rho=0.9$. To show how the cross-sectional correlations affect the connectedness measures, we compute the measures with an additional step in the estimation, considering only the diagonals of the covariance matrix of residuals and removing the cross-sectional dependence. In this way, we disentangle the influence of correlations from the true dynamics. In the text, we always present only the estimates on the simulated data and save the true values of measures in the Table~\ref{tab:trueparams} in the appendix \ref{app:tabsandfigs}.

\begin{table}[t]
	\tiny
	\centering
	\begin{tabular}{ccccccccccc}
	\toprule
		& & & \multicolumn{4}{c}{Connectedness}&\multicolumn{4}{c}{Connectedness without correlation}\\
		\cmidrule(r){4-7}
		\cmidrule(r){8-11}
		$ \beta $ & $ s $ & $ \rho $ & Total & $ (\pi/2, \pi) $ & $ (\pi/4, \pi/2) $ & $ (0, \pi/4) $ & Total & $ (\pi/2, \pi) $ & $ (\pi/4, \pi/2) $& $ (0, \pi/4) $ \\
	\midrule
0.00 & 0.00 & 0.00 & 0.18  &  0.19  &  0.19  &  0.19  &  0.09  &  0.10  &  0.10  &  0.10\\
     &      &      & (0.16) & (0.20) & (0.18) & (0.21) & (0.10) & (0.11) & (0.11) & (0.11)\\
0.00 & 0.00 & 0.90 & 44.68  &  44.75  &  44.74  &  44.72  &  0.47  &  0.41  &  0.41  &  0.41\\
     &      &      & (0.30) & (0.36) & (0.36) & (0.42) & (0.55) & (0.51) & (0.51) & (0.50)\\
0.90 & 0.09 & 0.00 & 37.65  &  0.69  &  1.26  &  37.77  &  37.24  &  0.64  &  1.22  &  37.84\\
     &      &      &  (4.55) & (0.76) & (0.76) & (4.36) & (4.64) & (0.93) & (0.92) & (4.81)\\
0.90 & 0.09 & 0.90 & 49.24  &  43.97  &  44.15  &  49.36  &  35.31  &  0.37  &  0.79  &  35.07\\
     &      &      & (0.33) & (0.64) & (0.62) & (0.27) & (6.27) & (0.45) & (0.53) & (5.10)\\
-0.90 & -0.09 & 0.00 &  39.33  &  38.91  &  0.96  &  0.87  &  38.97  &  38.68  &  0.81  &  0.71\\
     &      &       & (3.90) & (4.24) & (1.19) & (1.20) & (4.31) & (4.21) & (0.88) & (0.89)\\
-0.90 & -0.09 & 0.90 & 49.40  &  49.43  &  43.81  &  43.77  &  35.10  &  35.74  &  0.44  &  0.37\\
     &      &       & (0.32) & (0.23) & (0.62) & (0.63) & (6.28) & (4.95) & (0.30) & (0.29)\\
	\bottomrule
	\end{tabular}
	\caption{Simulation results. The first three columns describe the parameters for the simulation as described in Equation~(\ref{eq:VARsimulation}). We set $\beta=\beta_1=\beta_2$. The results are based on 100 simulations of VAR. The specified parameters have a length of 1000 and a burnout period of 100. The estimate is computed averaging over the 100 simulations, and the standard error is the sample standard deviation.}
	\label{tab:simulationVAR}
\end{table}

Table~\ref{tab:simulationVAR} shows the results. We can observe that the system's connectedness of two unconnected and uncorrelated processes is practically zero in both the total and all spectral parts. In case of correlated noises, the total connectedness with the estimated correlation matrix is estimated at approximately 45, with an equal footprint on all scales. Considering only diagonal elements from the estimated covariance matrix of the residuals and removing the cross-sectional dependence correctly estimates no connectedness at all frequencies.

For an AR coefficient equal to 0.9, the uncorrelated case shows that the connection between the processes is on the long-term part (as expected, due to the spectral density of the underlying process). However, introducing correlation increases the total connectedness and most of all obfuscates the source of the dynamics. Considering only the diagonals of the covariance matrix of the estimated residuals, we can see that the correlation in the estimated covariance matrix correctly exposes the underlying dynamics. The remaining case with the coefficient equal to -0.9 is very similar to the previous case, except the spectral mass is concentrated on the short frequencies. Otherwise, the qualitative results remain the same.

It is important to note that whereas coefficients with opposite signs of 0.9 and -0.9 generate the time series with equal connectedness, the source is from different parts of the spectra. This example motivates the usefulness of our measures, which are able to precisely locate the part of cross-spectra generating the connectedness.

Next, we move to the case where the two processes are not symmetric. With the simulation, we want to illustrate two important cases of how the connectedness arises. First, let us keep the parameter $s$ that governs the connection of the two processes through the lagged observation constant and change the spectral structure of the processes through the coefficient $\beta_2$.

\begin{table}[t]
	\tiny
	\centering
	\begin{tabular}{cccccccccccc}
	\toprule
		& & & & \multicolumn{4}{c}{Connectedness}&\multicolumn{4}{c}{Connectedness without correlation}\\
		\cmidrule(r){5-8}
		\cmidrule(r){9-12}
		$ \beta_1 $ & $ \beta_2 $ & $ s $ & $ \rho $ & Connectedness & $ (\pi/2, \pi) $ & $ (\pi/4, \pi/2) $ & $ (0, \pi/4) $ & Connectedness & $ (\pi/2, \pi) $ & $ (\pi/4, \pi/2) $& $ (0, \pi/4) $ \\
	\midrule
0.90 & 0.90 & 0.09 & 0.00 & 36.75  &  0.74  &  1.35  &  38.57  &  37.06  &  0.56  &  1.14  &  38.03\\
     &      &      &      & (4.79) & (1.32) & (1.27) & (3.94) & (4.64) & (0.45) & (0.49) & (4.24)\\
0.90 & 0.90 & 0.09 & 0.90 & 49.23  &  43.99  &  44.17  &  49.35  &  33.30  &  0.43  &  0.86  &  36.03\\
     &      &      &      & (0.32) & (0.51) & (0.49) & (0.24) & (5.90) & (0.46) & (0.53) & (4.80)\\
0.90 & 0.40 & 0.09 & 0.00 & 5.62  &  0.43  &  0.95  &  7.50  &  5.69  &  0.29  &  0.82  &  7.36\\
     &      &      &      & (1.52) & (0.31) & (0.32) & (2.19) & (1.48) & (0.07) & (0.19) & (1.94)\\
0.90 & 0.40 & 0.09 & 0.90 & 46.11  &  44.10  &  44.35  &  46.53  &  5.40  &  0.30  &  0.82  &  7.38\\
     &      &      &      & (0.36) & (0.57) & (0.53) & (0.30) & (1.62) & (0.06) & (0.17) & (2.17)\\
0.90 & 0.00 & 0.09 & 0.00 & 2.70  &  0.42  &  0.93  &  4.21  &  2.54  &  0.31  &  0.78  &  3.95\\
     &      &      &      & (0.91) & (0.38) & (0.35) & (1.16) & (0.68) & (0.08) & (0.21) & (1.23)\\
0.90 & 0.00 & 0.09 & 0.90 & 45.37  &  44.37  &  44.61  &  45.87  &  2.66  &  0.29  &  0.74  &  3.77\\
     &      &      &      & (0.36) & (0.42) & (0.39) & (0.37) & (0.78) & (0.06) & (0.16) & (1.20)\\
0.90 & -0.90 & 0.09 & 0.00 & 0.53  &  0.59  &  0.54  &  0.53  &  0.49  &  0.47  &  0.45  &  0.45\\
     &      &      &      & (0.15) & (0.38) & (0.20) & (0.31) & (0.10) & (0.12) & (0.10) & (0.11)\\
0.90 & -0.90 & 0.09 & 0.90 & 44.78  &  44.61  &  44.76  &  44.82  &  0.46  &  0.46  &  0.46  &  0.47\\
     &      &      &      & (0.30) & (0.36) & (0.35) & (0.35) & (0.09) & (0.06) & (0.11) & (0.15)\\
	\bottomrule
	\end{tabular}
	\caption{Simulation results. The first three columns describe the parameters for the simulation, as described in Equation~(\ref{eq:VARsimulation}). The results are based on 100 simulations of VAR with the specified parameters of length 1000 and a burnout period of 100. The estimate is computed by averaging over the 100 simulations, and the standard error is the simple sample standard deviation.}
	\label{tab:simulationVAR2}
\end{table}
The Table~\ref{tab:simulationVAR2} shows that in this case, the connectedness is arising due to the increase in the spectral similarity of the processes in question. This is an important example, as it shows that even a relatively small interaction at some frequency band can create strong connectedness among variables. Keeping the structure of the processes constant and increasing the parameter of interconnection increases the connectedness, as is documented by Table~\ref{tab:simulationVAR3} in the appendix \ref{app:tabsandfigs}.

The simulation exercise suggests possible sources of connectedness and motivates the usefulness of our measures. The role of covariance among the processes can be studied through the exclusion of the covariance terms, and the role of similarity can be examined through individual spectral densities; however, as mentioned, most of the economic series have similar spectral densities \citep{granger1966typical}. Our measures estimate the rich dynamics precisely.

\section{Systemic risk in the US financial sector}

In the past few decades, the literature has extensively studied the question of how financial firms are interconnected. Focusing on studying causality effects, co-movement, spillovers, connectedness, and systemic risk, researchers primarily try to answer the question using methods measuring the aggregate effects. Our interest is to measure frequency sources of volatility connectedness; hence, sources of systemic risk, as shocks to volatility, will impact future uncertainty differently. For example, fundamental changes in investors' expectations will have an impact on the market in the long term. These expectations are then transmitted to surrounding assets in portfolios differently than shocks with a short-term impact.

The early literature measuring the connectedness of markets in general was predominantly interested in contagion effects in market prices during crises. In a seminal paper, \citet{forbes2002no} show that if we account for volatilities of the price processes, the contagion effects disappear. This led to a rather strong statement of no-contagion, and interdependence among the markets remained the main effect of interest. However, as \citet{ross1989information} shows in his famous paper, volatility is the carrier of information in standard martingale price models. Hence, most of the later literature concentrates on the connectedness of volatilities. \citet{tse2002multivariate} concentrate on investigating the connection in the multivariate GARCH framework. They report high cross-correlations on the Forex market, national stock market, and Hang Seng sectoral indices. \citet{bae2003new} investigate the co-incidence of the extreme returns across markets and connect this measure by extreme value theory. They evaluate the contagion effects among various parts of the world, finding a high co-incidence of negative returns across markets. \cite{engle2012volatility} provide an exhaustive review of the empirical literature on volatility spillovers.

A broader picture concerning spillovers was later provided by \citet{diebold2009}, who explicitly investigated volatilities and returns separately and uncovered contagion effects in volatilities. In the same paper, the authors side-stepped the controversial topic of contagion, which had already been tied predominantly to financial crises and introduced the concept of spillovers, which refer to varying interdependency between the markets. Borrowing from both the contagion and interdependence notions, \citet{diebold2009} define a rigorous framework for measuring spillovers of returns and volatility across markets. These authors coin the term connectedness in their subsequent work \citep{diebold2011network}, in which they measure systemic risk in the US financial sector. Over the course of a few years, hundreds of studies in the literature successfully used this methodology to measure connectedness effects. However, the literature is still silent about the origins of the connectedness on business-cycle levels.

\subsection{Data}
\label{sec:data}

Considering volatility connectedness, we investigate the question of how market risk is connected at different frequencies. We study the intra-market connectedness of the US financial sector. We concentrate on the eleven major financial firms representing the financial sector of the US economy: Wells Fargo Co. (WFC), US Bancorp (USB), Morgan Stanley (MS), J.P. Morgan (JPM), Goldman Sachs (GS), Citibank (C), Bank of New York Mellon (BK), Bank of America (BAC), American Express (AXP), American International Group (AIG), and PNC Group (PNC). The dataset spans the years 2000 to 2016. We also investigate the connectedness of the system by adding Fannie Mae (FNM) and Freddie Mac (FRE) for the period ending in 2010. Because FNM and FRE went on the OTC market after 2010, the data are not publicly available, and the analysis cannot be performed on a longer time-span, although we argue that qualitative results for the overall connectedness are the same for smaller and larger systems. The complementary discussion is available in appendix \ref{app:fannie}.

For the computation of volatility, we restrict the analysis to daily logarithmic realized volatility, which is computed using 5-minute returns\footnote{Realized volatility for a given day is computed as the sum of squared intra-day returns.} during the business hours of the New York Stock Exchange, from 9:30 a.m. to 4:00 p.m. The data are time-synchronized by the same timestamps. We further eliminate transactions executed on Saturdays and Sundays, US federal holidays, December 24 to 26, and December 31 to January 2 because of the low activity on these days, which could lead to estimation bias. The data span the years 2000 to 2016, providing a sample of 4216 trading days. The period under study is informative in terms of market development, sentiment, and expectations, as we cover the 2007-2008 financial crisis and its aftermath years. The original raw data were obtained from TICK Data and www.price-data.com.

The descriptive statistics of the data can be found in the Table~\ref{tab:descstats} in the appendix \ref{app:tabsandfigs}.

\subsection{Time-frequency decomposition of US systemic risk}
\label{sec:intermarket_connectedness}

One of the issues that has recently gained importance in volatility modeling is giving up the assumption of global stationarity of the data \citep{stuaricua2005nonstationarities,engle2005spline} and focusing on local stationarity instead. When studying the connectedness of market risk using variance decompositions, it is important to face the nonstationarity of realized volatilities as zero frequency may dominate the rest of the frequencies when we study unconditional connectedness. The discussion gains importance when studying frequency dynamics because applying our measures blindly to the nonstationary data would result in false inferences.

Giving up the assumption of global stationarity of the data, we assume that the dynamics come from shifts in the unconditional variance of returns. This leads us to a convenient approximation of nonstationary data locally by stationary models. In essence, our approach is closely related to the one adopted by \cite{stuaricua2005nonstationarities}, although we study a multivariate system with quite different tools. We use the spectral representations of variance decompositions to recover the time-frequency dynamics of connectedness with a moving window of 300 trading days, where we confirm the stationarity of volatility. 

As our final model specification, we use a vector auto-regression with two lags, including a constant on the logarithm of volatilities to capture the dynamics in the window. We have experimented with different lag lengths with no material changes in the results. This only confirms the appropriateness of the approach because large changes in time-frequency dynamics due to different lags in the approximating VAR model would point to nonstationarities within windows, where a larger number of lags would approximate the information in the low frequencies.\footnote{In some sense, this analysis serves as a robustness check. We can make these results available upon request.} In large systems, a small-sample bias can occur; hence, we use a parametric bootstrap to obtain unbiased estimates of the connectedness measures, as suggested in \citep{engsted2014bias}.

\begin{figure}[t]
        \includegraphics[width=\textwidth,right]{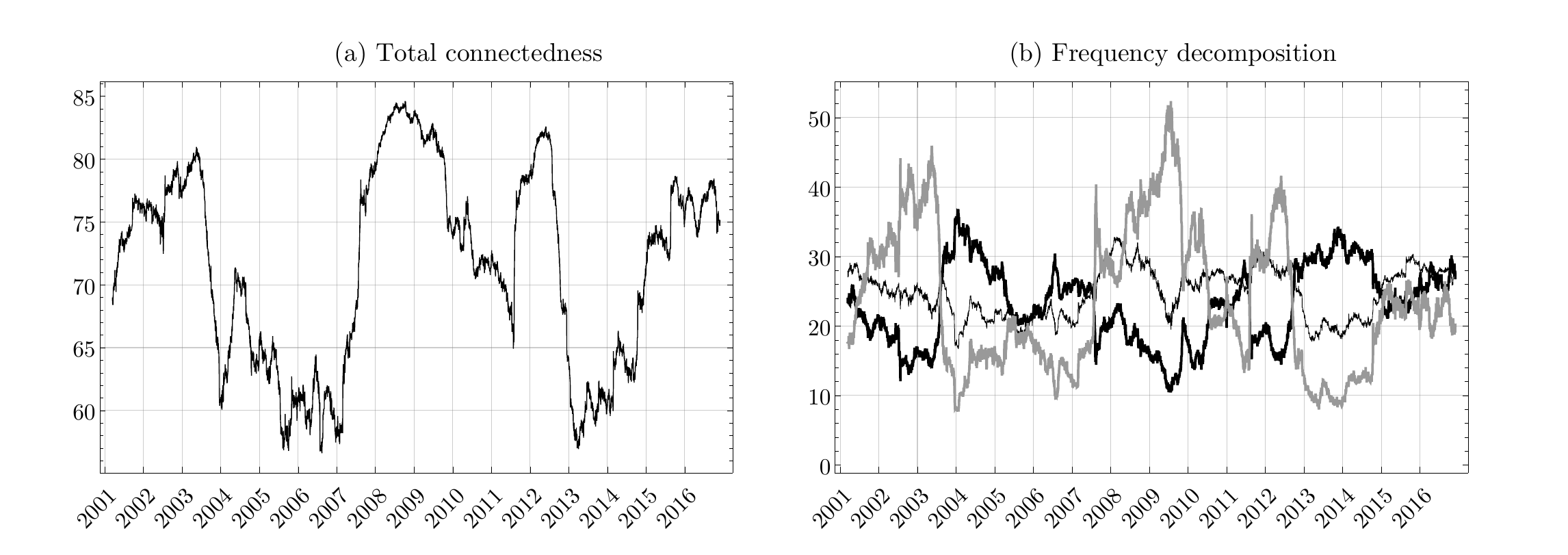}
    \caption{\footnotesize Dynamic frequency connectedness of the US financial sector. Plot (a) represents the total connectedness $\mathcal{C}$, computed on a moving window with a length of one year (300 days). Plot (b) represents the frequency connectedness $\mathcal{C}^{\mathcal{F}}_{d_s}$ with $d_1 \in [1,5]$ days in black bold, $d_2 \in (5,20]$ black, and $d_3 \in (20,300]$ gray bold. Note that all lines through the frequency bands $d_s$ sum to the total connectedness $\mathcal{C}$.}
    \label{fig:intermarket}
\end{figure}

Focusing on the locally stationary structure of the data, we study the time-frequency dynamics of connectedness. Figure~\ref{fig:intermarket} (a) reports the rich time dynamics of the total connectedness of system, as measured by time domain variance decompositions. Figure~\ref{fig:intermarket} (b) presents the decomposition of the total connectedness into frequency bands up to one week, one week to one month, and one month to 300 days, computed as $\mathcal{C}^{\mathcal{F}}_{d_s}$ on the bands corresponding to short-term ($d_1 \in [1,5]$), medium-term ($d_2 \in (5,20]$), and long-term ($d_3 \in (20,300]$). Note that the lowest frequency is bound at each time point by the window length.

Focusing first on the overall connectedness in Figure~\ref{fig:intermarket} (a), we can see that it ranges between 55\% and 85\%, with a substantial variation over the course of 16 years. Such a variation is expected because the studied period includes both calm and turbulent times in which shocks transmit across the system with different strengths. The overall connectedness bottoms during the calm times of 2005 and 2006, when shocks to uncertainty were transmitting less across the studied banks, thereby creating a less connected system. Connectedness peaks during the 2008 US financial crisis, when shocks created a large portion of future uncertainty and hence strong connectedness in the system. We also attribute the peak in connectedness to high uncertainty due to an insecure economic situation, in which banks were accused to be the main source of the economic instability. In this situation, shocks in the system created further uncertainty, which then transmitted across the system. As a result, banks were connected more strongly during the period following the crisis, including the beginning of 2012, when the European debt crisis peaked. Finally, connectedness starts declining after the ``\emph{Whatever it takes}'' speech by the ECB president Mario Draghi.\footnote{Our analysis of total connectedness suggests a picture that is qualitatively the same as \citet{diebold2011network}.}

Total connectedness is a useful measure that reveals how systemic risk changed over the studied period. It provides aggregate information over various economic cycles, and our main interest is to study the frequency sources of connectedness. In other words, Figure~\ref{fig:intermarket} (a) reveals that the total connectedness peaks during financial crises due to high uncertainty transmission, but it does not reveal whether shocks that create large connectivity in the system impact the system in the short term or in the long term. Since agents operate on different investment horizons, they may focus on different components of their consumption and in turn value assets through the expected utility from consumption with different persistence levels.\footnote{Asset pricing implications at different frequencies have been discussed in the literature by \cite{bandi2015business,dew2013asset}.} Hence, cyclical components will naturally generate shocks with heterogenous responses and thus various sources of connectedness, creating short-term, medium-term, and long-term systemic risk. 

Periods in which connectedness is created at high frequencies are periods when financial markets seem to process information rapidly, and a shock to one asset in the system mainly affects short-term cyclical behaviour (with responses at high frequencies). If the connectedness comes from the opposite part of the cross-spectral density, lower frequencies, it suggests that shocks are being transmitted for longer periods (with responses mainly at low frequencies). This behaviour may be attributed to fundamental changes in investors' expectations, which affect systemic risk in the longer term. These expectations are then transmitted to surrounding assets in portfolios. In a financial system where asset prices are driven by consumption growth with different cyclical components, shocks with heterogeneous responses create linkages with various degree of persistence and hence various sources of system-wide connectedness and systemic risk.

Figure~\ref{fig:intermarket} (b)\footnote{Complementary Figure~\ref{fig:bootabs} (a) in the appendix \ref{app:tabsandfigs} shows individual components with bootstrapped confidence bands .} reveals the frequency decomposition of connectedness, which supports our discussion as it shows rich time-frequency dynamics of connectedness. The first striking observation from the decomposition is that the periods of high total connectedness discussed above are driven mostly by low-frequency components ($d_3$ movements longer than one month). Hence, shocks creating uncertainty in the long term drive connectedness during the 2001 to mid-2003, mid-2007 to mid-2010, and 2012 periods. The intuition behind this is the overall high uncertainty about the financial system during these periods coupled with high uncertainty about the economic situation. This situation is accompanied by declining stock market prices with increasing volatility. The increasing uncertainty then translates into more persistent responses of investors to shocks. The observation of high connectedness of the system driven by low-frequency responses to shocks is then translated into the long-term uncertainty driving increasing systemic risk in these periods.

\begin{figure}[t]
        \includegraphics[width=\textwidth]{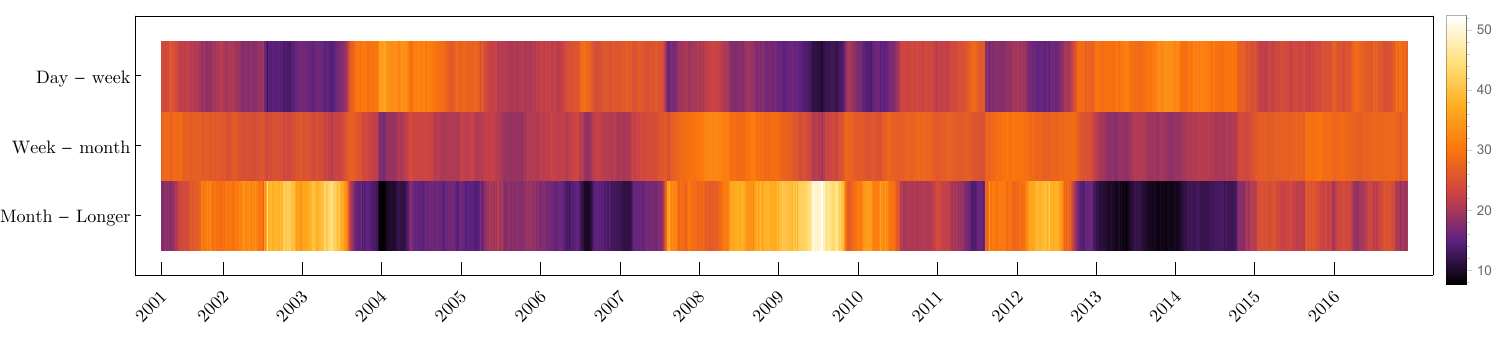}
    \caption{\footnotesize Time-frequency dynamics of the connectedness of the US financial sector. Frequency connectedness $\mathcal{C}^{\mathcal{F}}_{d_s}$ for $d_1 \in [1,5]$, $d_2 \in (5,20]$, and $d_3 \in (20,300]$ days representing day to week, week to month, and month to half a year are depicted on the vertical axis, and the horizontal axis shows time.}
    \label{fig:intermarket_heat}
\end{figure}

The structure of systemic risk dramatically changes during the rest of the sample period. We observe rich dynamics with short-term and medium-term components driving the connectedness. After the turbulent periods of high uncertainty, markets start to stabilize, and investors start to show less fear. The decreasing uncertainty in stable, growing markets translates into the fact that shocks creating future uncertainty in the system will transmit much faster, and their impact on the system will diminish after a few days, hence creating short-term connectedness. Figure~\ref{fig:intermarket} (b) documents the behaviour by increased short-term components after the consolidation from mid-2003. In addition, we document the rebound of short-term connectedness during 2010 before the European debt crisis peaks. The 2013-2014 period shows a strong influence of short-term components. This indicates that stock market participants were expecting that shocks to future uncertainty would have a short-term impact; hence, they were more certain about the long-term stability of the system. The overall systemic risk in this period was driven more by shocks with short-term responses. 

Interestingly, since the total connectedness increases from the beginning of 2015 until the end of the studied period, the increase is driven by all of the components. Hence, market participants process the shock responses homogeneously during this period. 

Figure~\ref{fig:intermarket_heat} contains the same information as Figure~\ref{fig:intermarket} (b) but shows the time-frequency dynamics from a different point of view, which serves as a helpful complementary visualization. In this figure, frequency bands form coloured ribbons, where the colour shows the strength of the connection, whereas the horizontal axis still holds time. One can view this representation as the three-dimensional space of connectedness at the time and frequency domains from the top, where the third axis showing the strength of connectedness at each time-frequency point is highlighted by colour. The heat map representation is useful because it more clearly visualizes the decomposition of the connectedness into time-frequency space. The interpretation is the same as Figure~\ref{fig:intermarket} (b) described in the previous paragraph. In addition, Figure~\ref{fig:bootabs} in the appendix \ref{app:tabsandfigs} shows a larger format picture with a $(5\%, 95\%)$ confidence band of the decomposition estimated by the bootstrap. The confidence band is sufficiently tight around the estimate to be informative. The confidence bands for other decompositions are also contained therein and are not described.

Our results indicate that when evaluating the systemic risk of a financial system, we should pay attention to short-, medium- and long-term linkages because they all play important roles in the system and tell us a different story of what is occurring. Before making further conclusions about the nature of the connectedness in US financial markets, we look deeper into its sources.

\begin{figure}[t]
        \includegraphics[width=\textwidth,right]{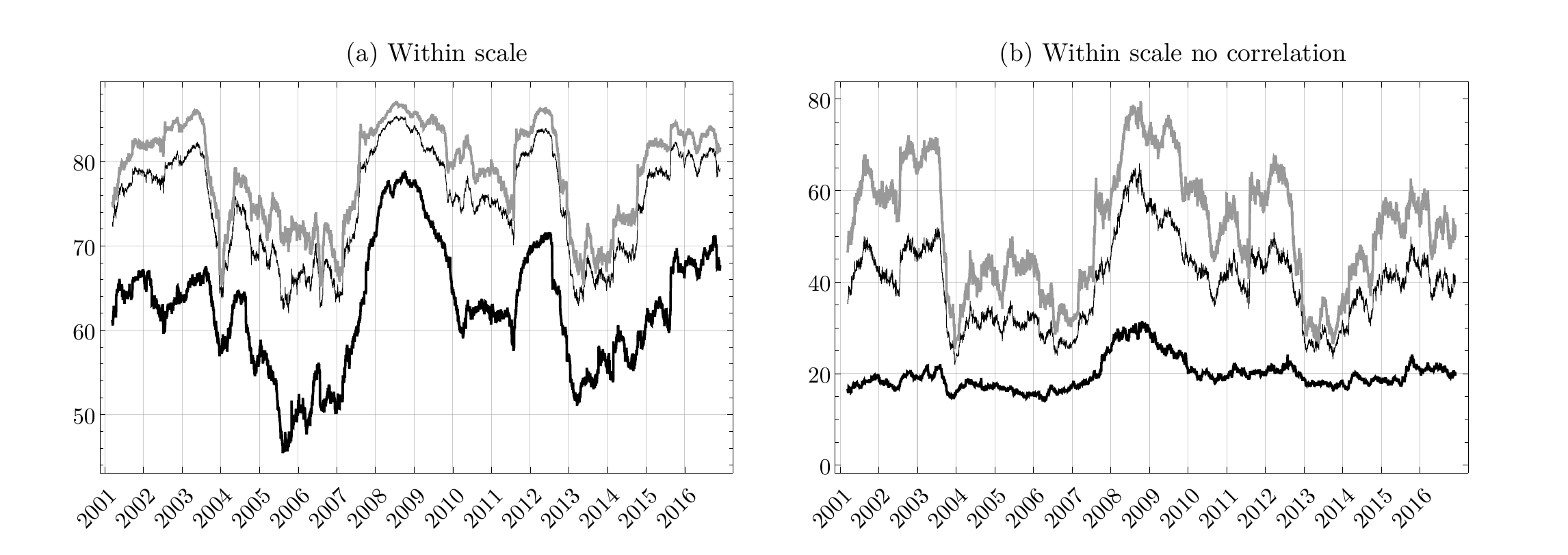}
    \caption{\footnotesize Dynamic within the connectedness of the US financial sector on frequency bands. Plot (a) presents the relative connectedness within the frequency band, $\mathcal{C}^{\mathcal{W}}_{d_s}$ with $d_1 \in [1,5]$ days in black bold, $d_2 \in (5,20]$ black, and $d_3 \in (20,300]$ gray bold lines. Plot (b) presents the relative connectedness within the frequency band without the effect of cross-sectional correlations.}
    \label{fig:intermarket_within}
\end{figure}

While the decomposition of the connectedness to frequency bands that always sums to total connectedness is the main interest, within connectedness serves as additional insight into the dynamics. Within connectedness shows how shocks transmit within frequency bands, and it is not weighted by the variance share at a given band. Ignoring information outside the considered band, connectedness within frequency bands can be understood as pure unweighted connectedness. Figure~\ref{fig:intermarket_within} (a) shows the within sectoral connectedness of the market. All frequencies share very similar time dynamics; hence, the rich time-frequency decomposition found in the previous part is mainly driven by the power of frequency responses, as should be expected.

The main reason why we look at the pure within connectedness is to study the effect of cross-sectional dependence on the connectedness. When using variance decompositions, we are mainly interested in finding causal effects, but these can be biased due to strong contemporaneous relationships. To determine whether there is a bias in the connectedness that we measure, we adjust the correlation matrix of VAR residuals by the cross-sectional correlations.

Figure~\ref{fig:intermarket_within} (b) shows within connectedness adjusted for this correlation effect. Strikingly, the structure changes dramatically, indicating that the high-frequency connectedness is mainly driven by cross-sectional correlations but that connectedness at lower frequencies is not affected as heavily, mainly during the crisis. One can infer that the increase in system connectedness during the crisis is mainly created by an increase in contemporaneous short-term correlations and causal longer-term connectedness. Hence, increased systemic risk has two main drivers. The first is an increase in contemporaneous correlations, and the second is an increase in the persistence of the shocks creating and transmitting long-term uncertainty.

\begin{figure}[t!]
    \centering
    \begin{subfigure}[b]{0.45\textwidth}
        \includegraphics[width=\textwidth,right]{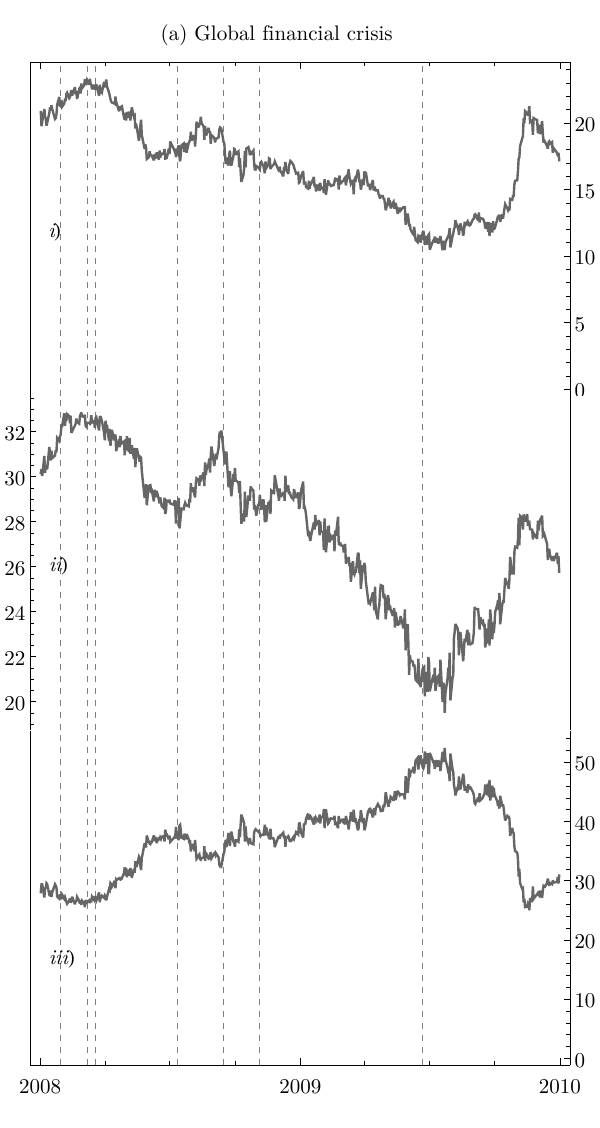}
    \end{subfigure}
    \begin{subfigure}[b]{0.45\textwidth}
        \includegraphics[width=\textwidth,right]{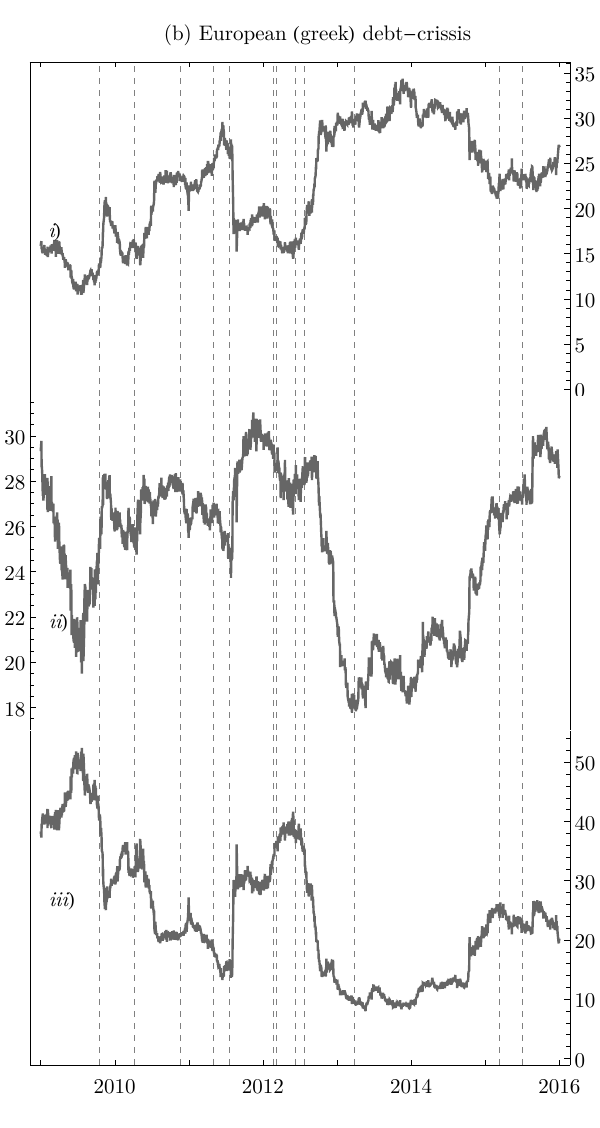}
    \end{subfigure}
    \caption{\footnotesize Decomposition of the influence of the main economic events in (a) the global financial crisis and (b) the European debt crisis on the connectedness measure. The individual lines represent connectedness measures at a given frequency band--more concretely: \emph{i)} connectedness from one day to one week, \emph{ii)} connectedness from one week to one month, and \emph{iii)} connectedness from one month to 300 days. The events in the European debt crisis, from left to right, are ``Papandreou reveals deficits'', ``Greece Activates 45bn EU/IMF Rescue Loans'', ``Irish bail-out'', ``Portuguese bail-out'', ``First debt write-down'', ``Second Greek bail-out'', ``Second debt write-down'', ``Spanish bank bail-out'', ``Whatever it takes'', ``Cypriot bail-out'', ``Start of ECB QE'', and ``Greek bail-out expires''. The events in the global financial crisis are ``FOMC lowered the Fed funds rate'', ``FED announces Term Auction Facility'', ``Bear Sterns buy-out, lowering FOMC rates'', ``IndyMac Bank Fails'', ``Lehman Brothers go bankrupt, next day FED buys AIG'', ``US General Election'', and ``World Bank projected that the global production for 2009 would fall by 2.9\%, the first decline since the second world war''.}
  \label{fig:crises}
\end{figure}

\subsubsection{Event study of evolution of frequency connectedness}

One of the main strengths of the connectedness framework of \citep{diebold2011network} is the possibility to \emph{time the events}, which means that important events usually track the evolution of connectedness. To manifest the strengths of our frequency decomposition, we focus on the event study of the two main periods with rich time-frequency dynamics: the global financial crisis of 2008-2010 and the European debt crisis of 2009-2016. The evolution of the frequency decomposition and timed events are shown in Figure~\ref{fig:crises}.

Let us start with the global financial crisis depicted by Figure~\ref{fig:crises} (a). The decomposition is dominated by the long-term component of connectedness that steadily increases until mid-2009, whereas the medium- and short-term connectedness steadily decrease. 

At the beginning of the period, short-term and medium-term connectedness were at the highest levels during the ``\emph{FOMC lowering the rates}'' or ``\emph{announcement of TAF}'' events. The market participants were interpreting that the shocks due to the two events would have a short-term impact. Hence, the shocks at the beginning of the financial crisis were viewed to have a short response and to create short-term connectedness.

The following surge in the long-term connectedness and the decline in short-term and medium-term connectedness can be attributed to changes in investors' beliefs about the stability of the system. More precisely, increasing uncertainty and the belief that the economic situation had deeper, systemic roots led to the period in which the shocks were transmitting with increasing persistence through the system. Surprisingly, the crash of ``\emph{IndyMac Bank}'' levelled off the development of connectedness for a while, changing investors' beliefs after the following crash of Lehman Brothers. 

From the downfall of Lehman Brothers until mid-2009, when it was already clear that the global economy was in its worst shape since the Second World War (as evidenced by the announcement by the World Bank in mid-2009 that global production would decline for the first time since that war), the evolution of frequency connectedness suggests the decreased importance of short-term connectedness and the increased importance of long-term connectedness. During this time, the shocks were impacting the system with increasing persistence, creating and transmitting uncertainty in the longer term. Hence, we document increasing uncertainty and insecurity about the studied system during this period. Overall, the global financial crisis appears to have been a continuously evolving crisis with no major surprises in that they were not reflected in the uncertainty and its transmission.

However, looking at Figure~\ref{fig:crises} (b), we see a story that is rather different from the global financial crisis. Starting with Greek PM Papandreou revealing deficits, the event has substantially moved the financial markets, instantly increasing the short-term connectedness by more than 5\% and the medium-term connectedness likewise. Considering that the short-term connectedness ranges from 10\% to 35\% throughout the sample, a 5\% jump is significant. Hence, investors were still considering that the shock would create short-term uncertainty.

The second Greek bail-out has changed the expectations on the markets, as the long-term connectedness has jumped by 15\%. The figure also provides evidence of the great success of Mario Draghi's speech in which he stated that the European Central Bank would do ``\emph{Whatever it took}'' to help Greece. 

After the speech, both long- and medium-term connectedness decreased to unprecedented levels, showing the increased confidence of investors accompanied by decreasing uncertainty. After the event, investors were more certain that shocks would have a less persistent impact on the system creating short-run connectedness.

\section{Conclusion} 
\label{sec:conclusion}

In this work, we contribute to the understanding of connectedness between economic variables by proposing to measure its frequency dynamics. Based on the spectral representations of variance decompositions and connectedness measures, we provide a general framework for disentangling the sources of connectedness between economic variables. Because shocks to economic activity have an impact on variables at different frequencies with different strengths, we view the frequency domain as a natural place for measuring the connectedness between economic variables.

As noted by \cite{diebold2009,diebold2012better,diebold2011network}, variance decompositions from approximating models are a convenient framework for empirical measurement of connectedness. A natural measure can be defined based on assessing shares of future uncertainty in one variable due to a shock arising in another variable in the system. Focusing on the frequency responses to shocks instead, we are interested in assessing shares of uncertainty in one variable due to shocks with different persistence levels. Moreover, we elaborate on the role of the correlation of the residuals in the magnitude of the connectedness.

In the empirical part, we investigate the connectedness of financial firms in the US, a powerful measure of systemic risk of the financial sector. We approximate the data locally and obtain rich time-frequency dynamics of connectedness. Economically, periods in which connectedness is created at high frequencies are periods when stock markets seem to process information rapidly and calmly, and a shock to one asset in the system will have an impact mainly in the short term. When the connectedness is created at lower frequencies, it suggests that shocks are persistent and are being transmitted for longer periods. The behaviour may be attributed to changes in investors' expectations, which impact the market in the longer term. The expectations are then transmitted to surrounding assets in portfolios. The two event studies of the global financial crisis during 2008 and the European debt crisis that followed in 2011 support our hypotheses. The results underline the importance of properly measuring the dynamics across time and frequencies and emphasize the important role of cross-sectional correlation in the connectedness origins.

The frequency-based approach opens fascinating new routes in understanding the connectedness of economic variables with important implications for the measurement of systemic risk. Further research applying our measures to wider areas of interest and different empirical modelling strategies will be important in uncovering the connectedness of assets within a market or industry and the connectedness across asset classes and international markets. It will also be important in providing grounds for further research in risk management, portfolio allocation, or business cycle analysis, where understanding the origins of connectedness is essential.

\newpage
{\footnotesize{
\setlength{\bibsep}{3pt}
\bibliographystyle{chicago}
\bibliography{BIBLIOGRAPHY}
}}

\newpage
\appendix
\section{Technical Appendix}
\subsection{Derivation of the GFEVD}
\label{app:genfevd}

Let us have the MA($\infty$) representation of the generalized VAR model (details in \citep{pesaran1998generalized,dees2007long}) given as
\begin{equation}
	\mathbf{x}_t = \boldsymbol \Psi(L) \boldsymbol \epsilon_t,
\end{equation}
with the covariance matrix of the errors $\boldsymbol \Sigma$. Because the errors are assumed to be serially uncorrelated, the total covariance matrix of the forecast error conditional at the information in $t-1$ is
\begin{equation}
	\boldsymbol \Omega_H = \sum_{h=0}^{H} \boldsymbol \Psi_h \boldsymbol \Sigma \boldsymbol \Psi'_h.
\end{equation}
Next, we define the covariance matrix of the forecast error conditional on knowledge of today's shock and future expected shocks to $j$-th equation. Starting from the conditional forecasting error,
\begin{equation}
	\boldsymbol \gamma_t^k(H) = \sum_{h=0}^{H} \boldsymbol \Psi_h \left[ \boldsymbol\epsilon_{t+H-h} - E(\boldsymbol \epsilon_{t+H-h} | \boldsymbol \epsilon_{k,t+H-h})   \right],
\end{equation} assuming normal distribution, we have
\begin{equation}
	\boldsymbol \gamma_t^k(H) = \sum_{h=0}^{H} \boldsymbol \Psi_h \left[ \boldsymbol \epsilon_{t+H-h} - \sigma_{kk}^{-1} \left(\boldsymbol \Sigma \right)_{\cdot k} \boldsymbol \epsilon_{k,t+H-h}  \right].
\end{equation}
Finally, the covariance matrix is
\begin{equation}
	\boldsymbol \Omega_H^k = \sum_{h=0}^{H} \boldsymbol \Psi_h \boldsymbol \Sigma \boldsymbol \Psi'_h - \sigma_{kk}^{-1} \sum_{h=0}^{H} \boldsymbol \Psi_h \left(\boldsymbol \Sigma \right)_{\cdot k} \left(\boldsymbol \Sigma \right)_{\cdot k}' \boldsymbol \Psi'_h.
\end{equation} Then, \begin{equation}
	\boldsymbol \Delta_{(j)kH} = (\boldsymbol \Omega_H - \boldsymbol \Omega_H^k)_{j,j} = \sigma_{kk}^{-1} \sum_{h = 0}^{H} \left( (\boldsymbol \Psi_h \boldsymbol \Sigma)_{j, k} \right)^2
\end{equation} is the unscaled $H$-step ahead forecast error variance of the $j$-th component with respect to the innovation in the $k$-th component. Scaling the equation yields the desired
\begin{equation}
	\left( \boldsymbol \theta_H \right)_{j,k} = \frac{\sigma_{kk}^{-1}\sum_{h = 0}^{H}\left( (\boldsymbol \Psi_h \boldsymbol \Sigma)_{j,k} \right)^2}{\sum_{h=0}^{H} (\boldsymbol \Psi_h \boldsymbol \Sigma \boldsymbol \Psi_h')_{j,j}}
\end{equation}

\subsection{Proofs}
\label{app:proofs}
\begin{proof}[Proposition~\ref{prop:1}]

	To prove the equality we need the following:
	\begin{align}
		\begin{split}
			\frac{1}{2 \pi} \int_{-\pi}^{\pi} \Gamma_j(\omega) \left(\bm{\mathfrak{f}} (\omega) \right)_{j,k} d \omega & = \frac{1}{2 \pi} \int_{-\pi}^{\pi} \frac{\left(\boldsymbol \Psi(e^{-i \omega}) \boldsymbol \Sigma \boldsymbol \Psi'(e^{+i\omega})\right)_{j,j}}{\frac{1}{2 \pi} \int_{-\pi}^{\pi} \left(\boldsymbol \Psi(e^{-i \lambda}) \boldsymbol \Sigma \boldsymbol \Psi'(e^{+i\lambda})\right)_{j,j} d \lambda} \frac{\sigma_{kk}^{-1} \left| \left( \boldsymbol \Psi \left( e^{-i \omega}  \right) \boldsymbol \Sigma \right)_{j,k} \right|^2 }{\left(\boldsymbol \Psi(e^{-i \omega}) \boldsymbol \Sigma \boldsymbol \Psi'(e^{+i\omega})\right)_{j,j}} d \omega\\
			& = \frac{1}{2 \pi} \int_{-\pi}^{\pi} \frac{\sigma_{kk}^{-1} \left| \left( \boldsymbol \Psi \left( e^{-i \omega}  \right) \boldsymbol \Sigma \right)_{j,k} \right|^2 }{\frac{1}{2 \pi} \int_{-\pi}^{\pi} \left(\boldsymbol \Psi(e^{-i \lambda}) \boldsymbol \Sigma \boldsymbol \Psi'(e^{+i\lambda})\right)_{j,j} d \lambda} d \omega \\
			& = \frac{\frac{1}{2 \pi} \int_{-\pi}^{\pi} \sigma_{kk}^{-1} \left| \left( \boldsymbol \Psi \left( e^{-i \omega}  \right) \boldsymbol \Sigma \right)_{j,k} \right|^2  d \omega}{\frac{1}{2 \pi} \int_{-\pi}^{\pi} \left(\boldsymbol \Psi(e^{-i \lambda}) \boldsymbol \Sigma \boldsymbol \Psi'(e^{+i\lambda})\right)_{j,j} d \lambda} \\
			& = \frac{\sigma_{kk}^{-1} \sum_{h=0}^{\infty} \left( \left( \boldsymbol \Psi_h \boldsymbol \Sigma \right)_{j,k} \right)^2}{\left( \sum_{h = 0}^{\infty} \left( \boldsymbol \Psi_h \boldsymbol \Sigma \boldsymbol \Psi'_h \right) \right)_{k,k}} \\
			& = \left( \boldsymbol \theta_{\infty} \right)_{j,k}
		\end{split}
	\end{align}
	Hence, the proof essentially simplifies to proving two equalities
	\begin{gather}
		\frac{1}{2 \pi} \int_{-\pi}^{\pi} \sigma_{kk}^{-1} \left| \left( \boldsymbol \Psi \left( e^{-i \omega}  \right) \boldsymbol \Sigma \right)_{j,k} \right|^2  d \omega = \sigma_{kk}^{-1} \sum_{h=0}^{\infty} \left( \left( \boldsymbol \Psi_h \boldsymbol \Sigma \right)_{j,k} \right)^2 \label{eq:enum} \\
		\frac{1}{2 \pi} \int_{-\pi}^{\pi} \left(\boldsymbol \Psi(e^{-i \lambda}) \boldsymbol \Sigma \boldsymbol \Psi'(e^{+i\lambda})\right)_{j,j} d \lambda = \left( \sum_{h = 0}^{\infty} \left( \boldsymbol \Psi_h \boldsymbol \Sigma \boldsymbol \Psi'_h \right) \right)_{k,k} \label{eq:denom}
	\end{gather}
	For the following steps we will leverage the standard integral
	\begin{equation}
		\frac{1}{2\pi} \int_{-\pi}^{\pi} e^{i \omega (u - v)} d \omega =
			\begin{cases}
				1 & \text{ for } u = v\\
				0 & \text{ for } u \neq v.
			\end{cases}
	\end{equation}

	This integral is mostly useful in cases when we have series $\sum_{h=0}^{\infty} \phi_{h} \psi_{h}$ and we want to arrive to spectral representation. Note that $\sum_{h=0}^{\infty} \phi_{h} \psi_{h} = \frac{1}{2\pi} \int_{-\pi}^{\pi} \sum_{v=0}^{\infty} \sum_{u=0}^{\infty} \phi_{u} \psi_{v} e^{i \omega (u - v)} d \omega$. Levering this knowledge we prove the Equation~(\ref{eq:enum})

	\begin{align}
		\begin{split}
			\sigma_{kk}^{-1} \sum_{h=0}^{\infty} \left( \left( \boldsymbol \Psi_h \boldsymbol \Sigma \right)_{j,k} \right)^2 &= \sigma_{kk}^{-1} \sum_{h=0}^{\infty} \left( \sum_{z=1}^{n} \left( \boldsymbol \Psi_h \right)_{j,z} \left( \boldsymbol \Sigma \right)_{z,k} \right)^2 \\
			& = \sigma_{kk}^{-1} \frac{1}{2\pi} \int_{-\pi}^{\pi} \sum_{u = 0}^{\infty} \sum_{v = 0}^{\infty} \left( \sum_{x = 1}^{n} \left( \boldsymbol \Psi_u \right)_{j,x} \left( \boldsymbol \Sigma \right)_{x,k} \right)\left( \sum_{y = 1}^{n} \left( \boldsymbol \Psi_v \right)_{j,y} \left( \boldsymbol \Sigma \right)_{y,k} \right) e^{i \omega (u-v)} d \omega\\
			& = \sigma_{kk}^{-1} \frac{1}{2\pi} \int_{-\pi}^{\pi} \sum_{u = 0}^{\infty} \sum_{v = 0}^{\infty} \left( \sum_{x = 1}^{n} \left( \boldsymbol \Psi_u e^{i \omega u} \right)_{j,x} \left( \boldsymbol \Sigma \right)_{x,k} \right)\left( \sum_{y = 1}^{n} \left( \boldsymbol \Psi_v e^{-i \omega v} \right)_{j,y} \left( \boldsymbol \Sigma \right)_{y,k} \right) d \omega\\
			& = \sigma_{kk}^{-1} \frac{1}{2\pi} \int_{-\pi}^{\pi} \left( \sum_{u = 0}^{\infty} \sum_{x = 1}^{n} \left( \boldsymbol \Psi_u e^{i \omega u} \right)_{j,x} \left( \boldsymbol \Sigma \right)_{x,k} \right)\left( \sum_{v = 0}^{\infty} \sum_{y = 1}^{n} \left( \boldsymbol \Psi_v e^{-i \omega v} \right)_{j,y} \left( \boldsymbol \Sigma \right)_{y,k} \right) d \omega \\
			& = \sigma_{kk}^{-1} \frac{1}{2\pi} \int_{-\pi}^{\pi} \left( \sum_{x = 1}^{n} \left( \boldsymbol \Psi \left(e^{i \omega} \right) \right)_{j,x} \left( \boldsymbol \Sigma \right)_{x,k} \right)\left( \sum_{y = 1}^{n} \left( \boldsymbol \Psi  \left( e^{-i \omega} \right) \right)_{j,y} \left( \boldsymbol \Sigma \right)_{y,k} \right) d \omega \\
			& = \sigma_{kk}^{-1} \frac{1}{2\pi} \int_{-\pi}^{\pi} \left( \left( \boldsymbol \Psi \left( e^{ - i \omega}  \right) \boldsymbol \Sigma \right)_{j,k} \right)\left( \left( \boldsymbol \Psi \left( e^{ i \omega}\right) \boldsymbol \Sigma \right)_{j,k} \right) d \omega \\
			& = \sigma_{kk}^{-1} \frac{1}{2\pi} \int_{-\pi}^{\pi} \left| \left( \boldsymbol \Psi \left( e^{-i \omega}\right) \boldsymbol \Sigma \right)_{j,k} \right|^2 d \omega
		\end{split}
	\end{align}

	We use the switch to the spectral representation of the MA coefficients in the second step. The rest is a manipulation with the last step invoking the definition of modulus squared of a complex number to be defined as $|z|^2 = z z^*$. Note that we can use this simplification without loss of generality, because the $MA(\infty)$ representation that is described by the coefficients $\boldsymbol \Psi_h$ has a spectrum that is always symmetric.

	Next, we concentrate on the Equation~(\ref{eq:denom}) levering similar steps and the positive semidefiniteness of the matrix $\boldsymbol \Sigma$ that ascertains that there exists $\boldsymbol P$ such that $\boldsymbol \Sigma = \boldsymbol P \boldsymbol P'.$

	\begin{align}
		\begin{split}
			\sum_{h = 0}^{\infty} \left( \boldsymbol \Psi_h \boldsymbol \Sigma \boldsymbol \Psi'_h \right) & = \sum_{h = 0}^{\infty} \left( \boldsymbol \Psi_h \boldsymbol P \right) \left( \boldsymbol \Psi_h \boldsymbol P \right)' \\
			& = \frac{1}{2\pi} \int_{-\pi}^{\pi} \sum_{u = 0}^{\infty} \sum_{v = 0}^{\infty} \left( \boldsymbol \Psi_u e^{i \omega u} \boldsymbol P \right) \left( \boldsymbol \Psi_v e^{-i \omega v} \boldsymbol P \right)' d \omega \\
			& = \frac{1}{2\pi} \int_{-\pi}^{\pi} \sum_{u = 0}^{\infty} \left( \boldsymbol \Psi_u e^{i \omega u} \boldsymbol P \right) \sum_{v = 0}^{\infty} \left( \boldsymbol \Psi_v e^{-i \omega v} \boldsymbol P \right)' d \omega \\
			& = \frac{1}{2\pi} \int_{-\pi}^{\pi} \left( \boldsymbol \Psi \left( e^{i \omega} \right) \boldsymbol P \right) \left( \boldsymbol \Psi \left( e^{-i \omega } \right) \boldsymbol P \right)' d \omega \\
			& = \frac{1}{2\pi} \int_{-\pi}^{\pi} \left( \boldsymbol \Psi \left( e^{i \omega} \right) \boldsymbol \Sigma  \boldsymbol \Psi' \left( e^{-i \omega } \right) \right) d \omega \\
		\end{split}
	\end{align}

	This completes the proof.
\end{proof}

\begin{proof}[Proposition~\ref{prop:2}]
	Using the Remark~\ref{rem:recomposition} and appropriate substitutions, we have:
	\begin{equation}
		\begin{split}
			\sum_{d_z \in D} \mathcal{C}^{\mathcal{F}}_{d_z} = \sum_{d_z \in D} \left(\frac{\sum  \widetilde{\boldsymbol \theta}_{d_z} }{\sum  \widetilde{\boldsymbol \theta}_{\infty} } - \frac{\text{Tr} \left\{\widetilde{\boldsymbol \theta}_{d_z} \right\}}{\sum  \widetilde{\boldsymbol \theta}_{\infty}}\right) = 1 - \frac{\sum_{d_z \in D} \text{Tr} \left\{\widetilde{\boldsymbol \theta}_{d_z} \right\}}{\sum \widetilde{\boldsymbol \theta}_{\infty} } = \\
			1 - \frac{\text{Tr} \left\{ \sum_{d_z \in D} \widetilde{\boldsymbol \theta}_{d_z} \right\} }{\sum \widetilde{\boldsymbol \theta}_{\infty} } = \mathcal{C}_{\infty},
		\end{split}
	\end{equation}
	where the next to last equality follows from the linearity of the trace operator.
\end{proof}

\begin{proof}[Proposition~\ref{prop:3}]
	Using the definition of the connectedness, we have
	\begin{gather}
		\mathcal{C}^{\mathcal{W}}_{(-\pi, \pi)} = \mathcal{C}_{\infty}\\
		\mathcal{C}^{\mathcal{F}}_{(-\pi, \pi)} = \frac{\left(\widetilde{\boldsymbol \theta}_{(-\pi,\pi)} \right)_{j,k}}{n} - \frac{\text{Tr} \left\{\widetilde{\boldsymbol \theta}_{\infty} \right\}}{\sum \widetilde{\boldsymbol \theta}_{\infty} } = \frac{n}{n} - \frac{\text{Tr} \left\{\widetilde{\boldsymbol \theta}_{\infty} \right\}}{\sum \widetilde{\boldsymbol \theta}_{\infty} } = 1 - \frac{\text{Tr} \left\{\widetilde{\boldsymbol \theta}_{\infty} \right\}}{\sum \widetilde{\boldsymbol \theta}_{\infty} } = \mathcal{C}_{\infty}
	\end{gather}
\end{proof}

\newpage
\clearpage
\section{Supplementary Tables and Figures} 

\subsection{Including Fannie Mae (FNM) and Freddie Mac (FRE)} 
\label{app:fannie}

\begin{figure}[h]
        \includegraphics[width=\textwidth]{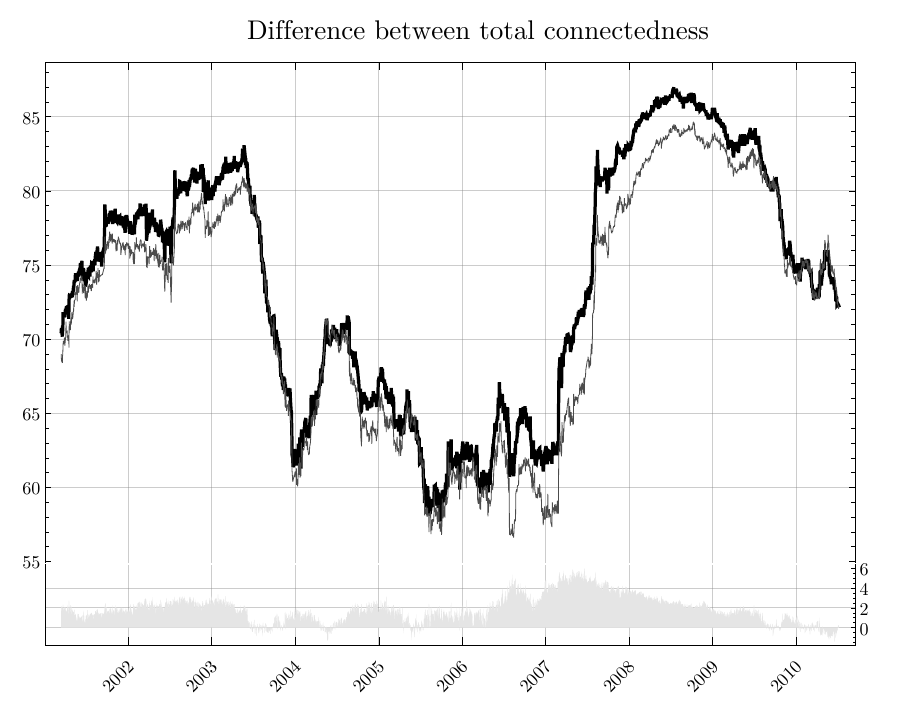}
    \caption{\footnotesize Time dynamics of connectedness of the US financial sector for system with and without FNM and FRE. The top part shows the level of connectedness and the bottom part shows the difference. The horizontal axis shows time. A positive difference indicates that the system including FNM and FRE is more connected than the system without those two stocks.}
    \label{fig:difference}
\end{figure}

As suggested in the data section, FRE and FNM were, for a substantial portion of the period, part of the US financial system. Hence, as a robustness measure, we compute the overall connectedness with the same specification over the shorter dataset when both FRE and FNM are publicly traded and compare the overall connectedness of the two systems to see how the results change. The result is shown in Figure~\ref{fig:difference}. We can see that most of the time, the system that includes the FRE and FNM was more connected than was the restricted system. The difference peaks at approximately 6\% in mid-2006, which is a relatively high number due to the nature of the connectedness, which is a mean of shares of variances created by shocks to other variables. Hence, including FRE and FNM in the system increases the systemic risk although the dynamics do not substantially differ from the analysis that omits FRE and FNM.

\newpage
\clearpage
\subsection{Supplementary Tables and Figures} 
\label{app:tabsandfigs}

\begin{table}[ht]
	\tiny
	\centering
	\begin{tabular}{cccccccccccc}
	\toprule
		& & & & \multicolumn{4}{c}{Connectedness as by DY12}&\multicolumn{4}{c}{Connectedness as by DY12, nullified correlation}\\
		\cmidrule(r){5-8}
		\cmidrule(r){9-12}
		$ \beta_1 $ & $ \beta_2 $ & $ s $ & $ \rho $ & Connectedness & $ (\pi/2, \pi) $ & $ (\pi/4, \pi/2) $ & $ (0, \pi/4) $ & Connectedness & $ (\pi/2, \pi) $ & $ (\pi/4, \pi/2) $& $ (0, \pi/4) $ \\
	\midrule
0.40 & -0.40 & 0.00 & 0.00 & 0.15  &  0.20  &  0.20  &  0.19  &  0.08  &  0.07  &  0.07  &  0.07\\
     &      &      &      & (0.12) & (0.21) & (0.19) & (0.21) & (0.08) & (0.07) & (0.07) & (0.07)\\
0.40 & -0.40 & 0.00 & 0.90 & 44.76  &  44.78  &  44.78  &  44.79  &  0.11  &  0.13  &  0.13  &  0.12\\
     &      &      &      & (0.35) & (0.35) & (0.35) & (0.36) & (0.11) & (0.18) & (0.17) & (0.17)\\
0.40 & -0.40 & 0.20 & 0.00 & 3.37  &  3.49  &  3.30  &  3.20  &  3.48  &  3.44  &  3.47  &  3.49\\
     &      &      &      & (0.60) & (0.99) & (0.78) & (1.08) & (0.64) & (0.67) & (0.68) & (0.69)\\
0.40 & -0.40 & 0.20 & 0.90 & 45.03  &  43.88  &  45.49  &  45.98  &  3.53  &  3.55  &  3.55  &  3.55\\
     &      &      &      & (0.34) & (0.48) & (0.32) & (0.31) & (0.60) & (0.54) & (0.62) & (0.67)\\
0.40 & -0.40 & 0.59 & 0.00 & 23.11  &  23.07  &  22.86  &  22.75  &  23.14  &  22.93  &  22.93  &  22.97\\
     &      &      &      & (0.93) & (1.55) & (1.17) & (1.90) & (0.96) & (1.18) & (1.00) & (1.27)\\
0.40 & -0.40 & 0.59 & 0.90 & 46.86  &  41.53  &  47.75  &  48.43  &  24.38  &  23.00  &  22.80  &  22.74\\
     &      &      &      & (0.29) & (0.70) & (0.17) & (0.16) & (0.97) & (0.90) & (1.14) & (1.45)\\
0.40 & -0.40 & -0.20 & 0.00 & 3.48  &  3.55  &  3.49  &  3.46  &  3.42  &  3.38  &  3.35  &  3.34\\
     &      &      &      & (0.64) & (0.89) & (0.80) & (1.05) & (0.65) & (0.69) & (0.67) & (0.68)\\
0.40 & -0.40 & -0.20 & 0.90 & 45.03  &  45.84  &  44.45  &  43.45  &  3.56  &  3.43  &  3.41  &  3.40\\
     &      &      &      & (0.31) & (0.28) & (0.37) & (0.46) & (0.61) & (0.65) & (0.57) & (0.54)\\
0.40 & -0.40 & -0.59 & 0.00 & 23.07  &  23.25  &  22.98  &  22.83  &  23.16  &  23.23  &  23.15  &  23.16\\
     &      &      &      & (0.94) & (1.49) & (1.08) & (1.68) & (1.04) & (1.30) & (0.95) & (1.25)\\
0.40 & -0.40 & -0.59 & 0.90 & 46.88  &  48.31  &  45.31  &  39.56  &  24.70  &  23.07  &  23.09  &  23.13\\
     &      &      &      & (0.29) & (0.17) & (0.34) & (0.78) & (0.92) & (1.56) & (0.86) & (0.85)\\
	\bottomrule
	\end{tabular}
	\caption{Simulation results. The first three columns describe parameters for the simulation as described in Equation~(\ref{eq:VARsimulation}). The results are based on 100 simulations of VAR with the specified parameters of length 1000 and a burnout period of 100. The estimate is computed by average over the 100 simulation, and the standard error is the simple sample standard deviation. The numbers are multiplied by hundred.}
	\label{tab:simulationVAR3}
\end{table}

\begin{table}
\tiny
\centering
\begin{tabular}{cccccccc}
\toprule
$\beta_1$ & $\beta_2$ & s & $\rho$ & Connectedness & $(\pi/2, \pi)$ & $(\pi/4, \pi/2)$ & $(0, \pi/4)$\\
\midrule
0.0 & 0.0 & 0.00 & 0.0 & 0.00 & 0.00 & 0.00 & 0.00\\
0.0 & 0.0 & 0.00 & 0.9 & 44.75 & 44.75 & 44.75 & 44.75\\
0.9 & 0.9 & 0.09 & 0.0 & 40.50 & 0.30 & 0.90 & 41.15\\
0.9 & 0.9 & 0.09 & 0.9 & 49.47 & 44.25 & 44.41 & 49.51\\
-0.9 & -0.9 & 0.09 & 0.0 & 40.50 & 40.77 & 0.34 & 0.24\\
\addlinespace
-0.9 & -0.9 & 0.09 & 0.9 & 41.28 & 41.01 & 45.22 & 45.22\\
0.9 & 0.4 & 0.09 & 0.0 & 5.66 & 0.32 & 0.88 & 7.48\\
0.9 & 0.4 & 0.09 & 0.9 & 46.09 & 44.25 & 44.48 & 46.56\\
0.9 & 0.0 & 0.09 & 0.0 & 2.59 & 0.32 & 0.80 & 3.97\\
0.9 & 0.0 & 0.09 & 0.9 & 45.40 & 44.25 & 44.51 & 45.98\\
\addlinespace
0.9 & -0.9 & 0.09 & 0.0 & 0.45 & 0.45 & 0.45 & 0.45\\
0.9 & -0.9 & 0.09 & 0.9 & 44.76 & 44.26 & 44.97 & 45.26\\
0.4 & -0.4 & 0.00 & 0.0 & 0.00 & 0.00 & 0.00 & 0.00\\
0.4 & -0.4 & 0.00 & 0.9 & 44.75 & 44.75 & 44.75 & 44.75\\
0.4 & -0.4 & 0.20 & 0.0 & 3.33 & 3.33 & 3.33 & 3.33\\
\addlinespace
0.4 & -0.4 & 0.20 & 0.9 & 45.01 & 43.52 & 45.62 & 46.28\\
0.4 & -0.4 & 0.59 & 0.0 & 23.08 & 23.08 & 23.08 & 23.08\\
0.4 & -0.4 & 0.59 & 0.9 & 46.87 & 40.94 & 47.86 & 48.64\\
0.4 & -0.4 & -0.20 & 0.0 & 3.33 & 3.33 & 3.33 & 3.33\\
0.4 & -0.4 & -0.20 & 0.9 & 45.01 & 46.05 & 44.27 & 43.00\\
\addlinespace
0.4 & -0.4 & -0.59 & 0.0 & 23.08 & 23.08 & 23.08 & 23.08\\
0.4 & -0.4 & -0.59 & 0.9 & 46.87 & 48.51 & 45.13 & 38.84\\
\bottomrule
\end{tabular}
\caption{The true values for connectedness in the VAR settings}
\label{tab:trueparams}
\end{table}

\begin{table}
\tiny
\centering
    \begin{tabular}{llrrrrrrll}
    \toprule
    Institution & Ticker & No. of obs. & Mean & Median & St. dev. & Skewness & Kurtosis & Start date & End date\\
    \midrule
    AIG & AIG & 4216 & 2.14 & 1.46 & 2.82 & 10.17 & 202.03 & 2000-01-03 & 2016-11-30\\
    American Express & AXP & 4216 & 1.63 & 1.30 & 1.15 & 2.88 & 18.52 & 2000-01-03 & 2016-11-30\\
    Bank of America & BAC & 4216 & 1.76 & 1.36 & 1.84 & 14.12 & 448.59 & 2000-01-03 & 2016-11-30\\
    Bank of New York Mellon & BK & 4216 & 1.72 & 1.34 & 1.41 & 6.53 & 88.15 & 2000-01-03 & 2016-11-30\\
    Citigroup & C & 4216 & 2.11 & 1.47 & 12.85 & 62.32 & 3985.03 & 2000-01-03 & 2016-11-30\\
    \addlinespace
    Fannie Mae & FNM & 2617 & 3.11 & 1.52 & 4.19 & 4.82 & 40.03 & 2000-01-03 & 2010-07-07\\
    Freddie Mac & FRE & 2617 & 3.16 & 1.47 & 4.58 & 5.55 & 55.22 & 2000-01-03 & 2010-07-07\\
    Goldman Sachs & GS & 4216 & 1.70 & 1.36 & 1.13 & 4.26 & 38.42 & 2000-01-03 & 2016-11-30\\
    J.P. Morgan & JPM & 4216 & 1.74 & 1.41 & 1.21 & 3.19 & 20.82 & 2000-01-03 & 2016-11-30\\
    Morgan Stanley & MS & 4216 & 2.18 & 1.76 & 1.80 & 7.39 & 104.67 & 2000-01-03 & 2016-11-30\\
    \addlinespace
    PNC Group & PNC & 4216 & 1.58 & 1.19 & 1.32 & 4.21 & 33.57 & 2000-01-03 & 2016-11-30\\
    US Bancorp & USB & 4216 & 1.62 & 1.23 & 1.25 & 3.32 & 22.18 & 2000-01-03 & 2016-11-30\\
    Wells Fargo & WFC & 4216 & 1.62 & 1.25 & 1.67 & 17.55 & 653.86 & 2000-01-03 & 2016-11-30\\
    \bottomrule
    \end{tabular}

\caption{The descriptive statistics of the volatility data}
\label{tab:descstats}
\end{table}

\begin{figure}[t]
    \centering
    \begin{subfigure}[b]{0.45\textwidth}
        \includegraphics[width=\textwidth,right]{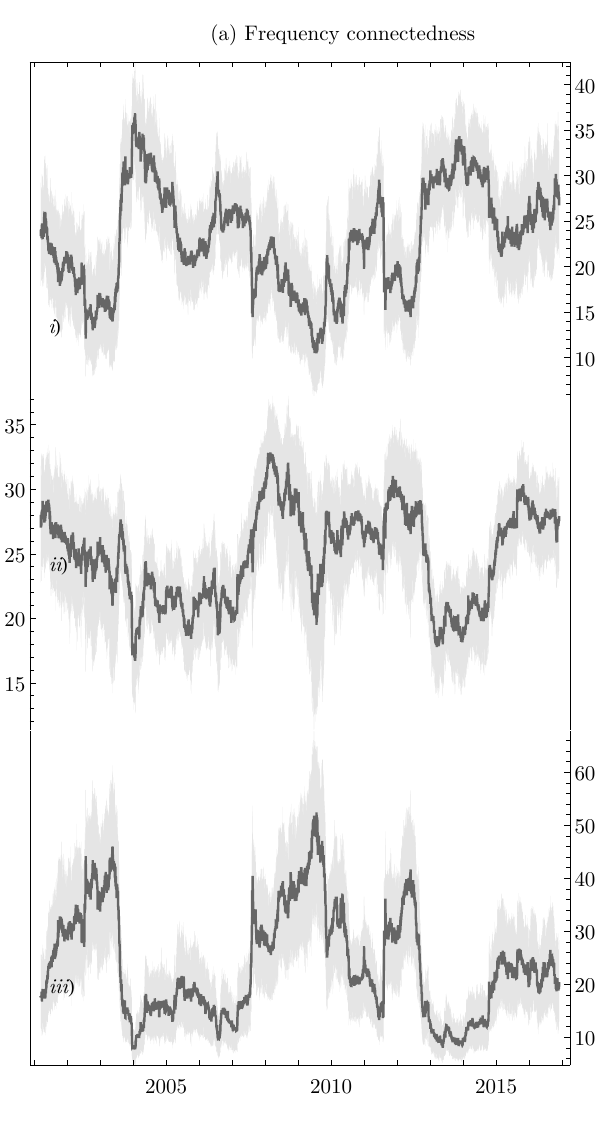}
    \end{subfigure}
    \begin{subfigure}[b]{0.45\textwidth}
        \includegraphics[width=\textwidth,right]{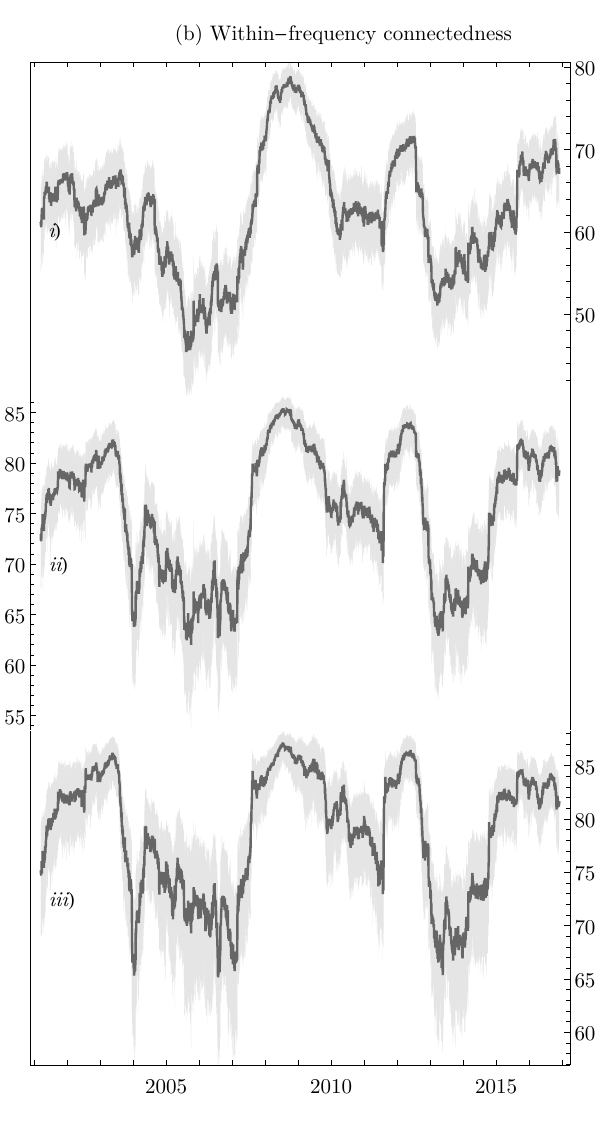}
    \end{subfigure}
    \caption{The decomposition of connectedness with cross-sectional dependence. The individual lines represent connectedness measures at a given frequency band, more concretely: \emph{i)} connectedness from one day to one week, \emph{ii)} connectedness from one week to one month, and \emph{iii)} connectedness from one month to 300 days. The shaded area represents the space between the 5\% and 95\% quantiles of the bootstrapped measure.}
  \label{fig:bootabs}
\end{figure}

\begin{figure}[t]
    \centering
    \begin{subfigure}[b]{0.45\textwidth}
        \includegraphics[width=\textwidth,right]{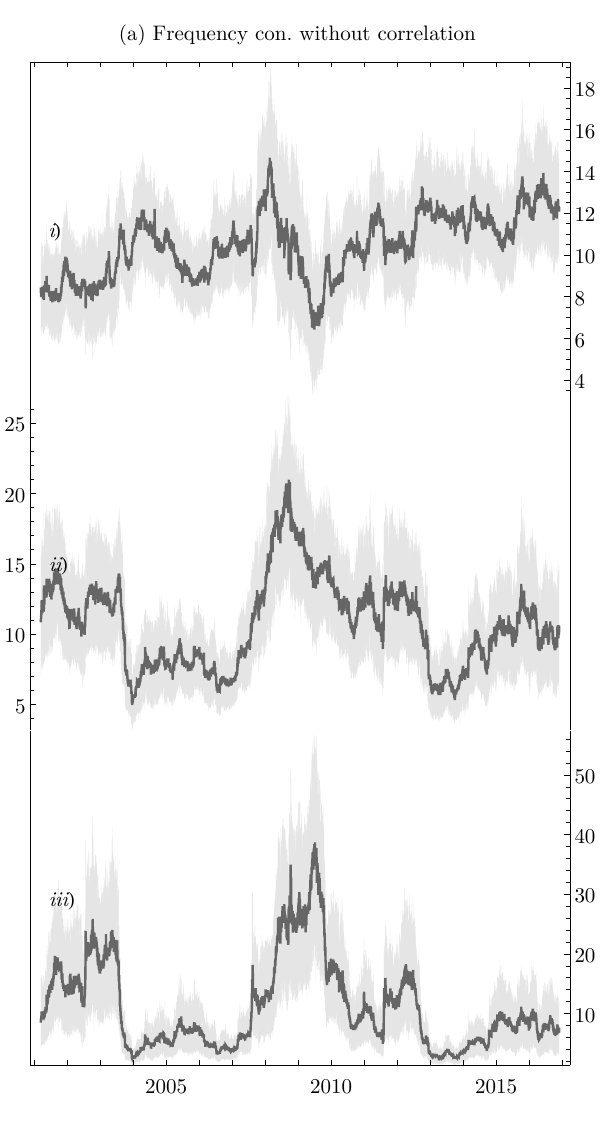}
    \end{subfigure}
    \begin{subfigure}[b]{0.45\textwidth}
        \includegraphics[width=\textwidth,right]{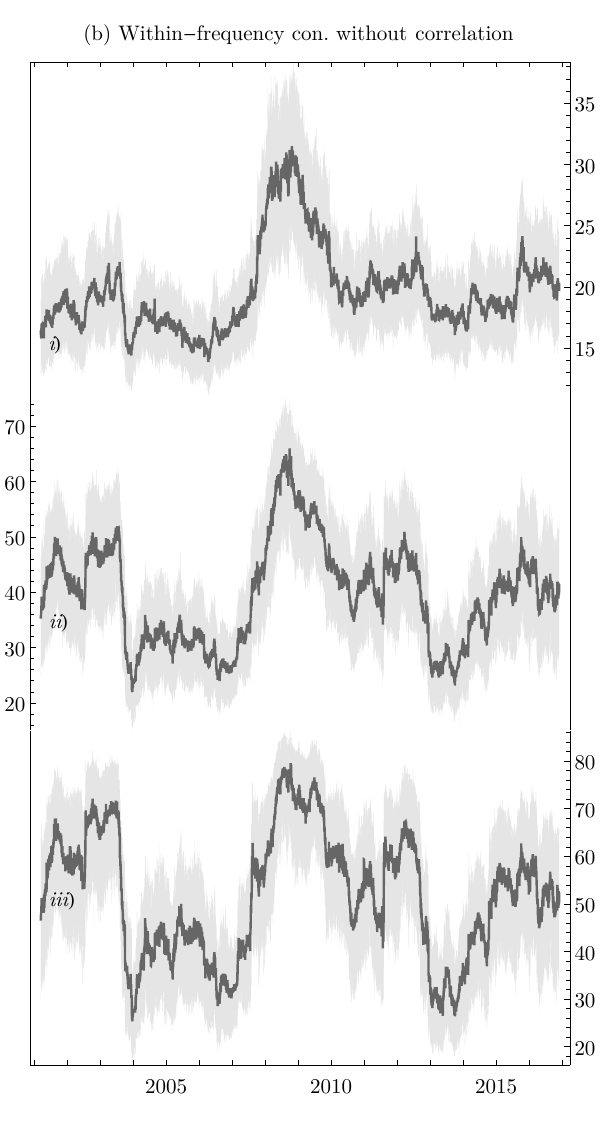}
    \end{subfigure}
    \caption{The decomposition of connectedness \emph{without} cross-sectional dependence. The individual lines represent connectedness measures at a given frequency band, more concretely: \emph{i)} connectedness from one day to one week, \emph{ii)} connectedness from one week to one month, and \emph{iii)} connectedness from one month to 300 days. The shaded area represents the space between the 5\% and 95\% quantiles of the bootstrapped measure.}
  \label{fig:bootwithout}
\end{figure}
\end{document}